\documentclass[12pt,a4paper]{article}
\usepackage{latexsym,amssymb,epsfig,rotating}

\newcommand{\bm}[1]{\mbox{\boldmath$#1$}}
\def\0{\phantom{0}}

\begin{document}
\pagenumbering{arabic}
\baselineskip25pt

\begin{center}
{\bf \large Comprehensive study of the vapour-liquid equilibria of the pure two-centre
Lennard-Jones plus pointdipole fluid} \\
\bigskip
\renewcommand{\thefootnote}{\fnsymbol{footnote}}
J\"urgen Stoll, Jadran Vrabec\footnote[1]{author for correspondence, Tel.: ++49-711/685-6107, Fax: ++49-711/685-7657, Email: vrabec@itt.uni-stuttgart.de }, Hans Hasse \\
\renewcommand{\thefootnote}{\arabic{footnote}}
Institut f\"ur Technische Thermodynamik und Thermische Verfahrenstechnik, \\
Universit\"at Stuttgart, D-70550 Stuttgart, Germany \\
\end{center}
\begin{abstract}
\baselineskip25pt
Results of a systematic investigation of the vapour-liquid equilibria of 38 individual two-centre Lennard-Jones plus axial pointdipole model fluids (2CLJD) are reported over a range of reduced dipolar momentum $0 \le \mu^{*2} \le 20$ and of reduced elongation $0 \le L^* \le 1.0$. Temperatures investigated are from about 55 \% to about 95 \% of the critical temperature of each fluid. The $N\!pT$+Test Particle Method is used for the generation of vapour pressures, saturated densities, and saturated enthalpies. For the lowest temperatures, these data are calculated with highly accurate chemical potentials obtained from the gradual insertion method. Critical temperatures $T^*_{\rm c}$ and densities $\rho^*_{\rm c}$ are obtained from Guggenheim's equations. Empirical correlations for critical data $T^*_{\rm c}$ and $\rho^*_{\rm c}$ as well as for saturated densities $\rho'^*$, $\rho''^*$, and  vapour pressures $p^*_{\sigma}$ are developed as global functions of the model parameters. They describe the simulation data generally within their statistical uncertainties. Critical pressures and acentric factors of the 2CLJD fluid can be calculated from these correlations. The present results are a sound basis for adjustments of the model parameters $\mu^{*2}$, $L^*$, $\sigma$, and $\epsilon$ to experimental VLE data of real fluids.
\end{abstract}

\section{Introduction}

Knowledge of vapor-liquid equilibria (VLE) is important in many problems in engineering and natural sciences. Among the different ways to model vapor-liquid equilibria, molecular simulation has the highest potential to yield significant improvements compared to existing models, especially in terms of predictive power. However, further efforts are needed until molecular simulation based models and tools will be sufficiently developed so that their advantages can help process engineers in their real world tasks. 

One of the main problems to be overcome is the lack of intermolecular interaction models describing vapor-liquid equilibria of real fluids with technically relevant accuracy. As in most applications mixtures are of interest, it is necessary to model pure fluids with a family of compatible interaction models, allowing their application for mixtures using simple combining rules.

Various Stockmayer fluids, i.e. one-centre Lennard-Jones plus pointdipole (1CLJD) fluids, and two-centre Lennard-Jones plus pointdipole (2CLJD) fluids have been studied by other authors. These model fluids have been applied to a number of real dipolar fluids. Stockmayer parameters are presented by van Leeuwen \cite{leeuwen941} for various real substances and by Gao et al. \cite{gao9787} for alternative refrigerants. 2CLJD parameters for various refrigerants are reported by Vega et al. \cite{vega8910}, Kohler and van Nhu \cite{kohler9379}, Kriebel et al. \cite{kriebel9815} and by L\'{\i}sal et al. \cite{lisal9916}. Model parameters for the dipolar Kihara potential, which is similar to the 2CLJD potential model, are given by Lago et al. \cite{lago9767} for organic solvents.

The search for an appropriate interaction model for a given fluid is usually a time consuming process. In general, the focus lies on {\it one pure fluid}, where the optimization of the potential model is done by a number of simulations with subsequent variation of the model parameters, cf., e.g., van Leeuwen et al. \cite{leeuwen9518} for methanol.

In a previous work we followed a new route to develop quantitative interaction models that allows fast adjustments of model parameters to experimental data for a given {\it class of pure fluids}. For the two-centre Lennard-Jones plus pointquadrupole (2CLJQ) fluid systematic studies of the vapour-liquid equilibria were carried out. The results were correlated as a function of the model parameters of the 2CLJQ fluid \cite{stoll0133}. Using these correlations, it was possible to determine the model parameters for a large number of real quadrupolar fluids \cite{vrabec0112}. It has been shown, that these molecular models can successfully be applied for the description of vapour-liquid equilibria of binary and multicomponent mixtures \cite{stoll02}.

Encouraged by these favorable results, in this work an analogous investigation is carried out for the important class of dipolar fluids. Only a few studies of VLE of the 2CLJD fluid are available in the literature \cite{lisal9916,lisal0036,lisal9719,dubey9399}, which however cover only a comparatively narrow range of the model parameters. Furthermore, VLE of some submodels of the 2CLJD fluid have previously been investigated: Galassi and Tildesley \cite{galassi9411}, Kriebel et al. \cite{kriebel9538}, and Kronome et al. \cite{kronome9827} the 2CLJ, van Leeuwen \cite{leeuwen941}, van Leeuwen et al. \cite{leeuwen9327}, Gao et al. \cite{gao9787}, Garz\'on et al. \cite{garzon9436}, and Smit et al. \cite{smit8976} the Stockmayer fluid (1CLJD). The simulation techniques applied were Gibbs-Duhem integration or the Gibbs Ensemble Monte Carlo method. The idea followed in this work is to study the VLE of the 2CLJD model fluid systematically and in detail over the whole relevant range of model parameters. Using reduced coordinates, for the symmetric 2CLJD fluid, only two parameters have to be varied: the dimensionless LJ centre-centre distance $L^*$ and the dimensionless squared dipolar momentum $\mu^{*2}$. The parameter space of interest can be covered with acceptable accuracy by studying 38 individual 2CLJD fluids with different values for $L^*$ and $\mu^{*2}$. The simulation results are correlated empirically in order to be able to interpolate between the discrete pairs of $L^*$ and $\mu^{*2}$.

Given the results from the present study, it is straightforward to adjust the molecular interaction parameters of the 2CLJD fluid to experimental VLE data of real dipolar fluids. Properties like the critical values of temperature and pressure, the acentric factor or the saturated liquid density and vapour pressure are available as functions of the molecular interaction parameters. Hence, the development of the molecular interaction model for a given substance is not more difficult than the adjustment of parameters of phenomenological thermodynamic models. 

This work covers the basic molecular simulations for 38 individual 2CLJD fluids and the development of the correlations together with a discussion of these results. The application to real fluids will be presented separately. 

For the calculation of vapour-liquid equilibria the $N\!pT$+Test Particle Method \cite{moeller9046} was chosen due to favourable experience with that method especially concerning accuracy.
\clearpage

\section{Investigated model class}

In this work pure two-centre Lennard-Jones plus axial pointdipole (2CLJD) fluids are studied. The 2CLJD potential model is composed of two identical Lennard-Jones sites a distance $L$ apart (2CLJ), forming the molecular axis, plus a pointdipole (D) $\bm \mu$ placed in the geometric centre of the molecule along the molecular axis. The pair potential $u_{\rm 2CLJD}$ and further technical details are described in the Appendix A.

The parameters $\sigma$ and $\epsilon$ of the 2CLJD pair potential were used for the reduction of the thermodynamic properties as well as the model parameters $L$ and $\mu^2$: $T^*=Tk/\epsilon$, $p^*=p\sigma^3/\epsilon$, $\rho^*=\rho\sigma^3$, $h^*=h/\epsilon$, $L^*=L/\sigma$, $\mu^{*2}=\mu^2/\left( 4 \pi \epsilon_0 \cdot \epsilon \sigma^3 \right)$, $\Delta t^*=\Delta t\sqrt{m/\epsilon}/\sigma$.

The reduced parameters $L^*$ and $\mu^{*2}$ were varied in this investigation: $L^*=0$; $0.2$; $0.4$; $0.505$; $0.6$; $0.8$; $1.0$ and $\mu^{*2}=0$; $3$; $6$; $9$; $12$. All combinations of these values lead to a set of 35 model fluids. Additionally to these, the fluids with $L^*=0.0$; $0.2$ and $\mu^{*2}=16$ as well as $L^*=0.0$ and $\mu^{*2}=20$ were investigated, as very strong dipolar momenta are only realistic for molecules with small elongation $L^*$. So, altogether 38 model fluids are considered.
\clearpage

\section{Molecular simulation method for VLE data}

For all 38 model fluids the $N\!pT$+Test Particle Method ($N\!pT$+TP Method) proposed by M\"oller and Fischer \cite{moeller9046,moeller9214,moeller9435} was applied to obtain the VLE data. The $N\!pT$+TP Method performs separate $N\!pT$ simulations in the liquid and the vapour phase and uses information on the chemical potentials to calculate the VLE. Vapour pressures, saturated densities and residual enthalpies $h^{\mbox{\scriptsize res*}} \left( T^*,\rho^* \right) =h^* \left( T^*,\rho^* \right)-h^{\mbox{\scriptsize id*}} \left( T^* \right)$ in equilibrium are evaluated.

Test particle insertion for the calculation of the residual chemical potential $\mu^{\mbox{\scriptsize res*}}$ in the $N\!pT$ ensemble is based in general on Widom's method \cite{widom6328}, which, however, yields $\mu^{\mbox{\scriptsize res*}}$ with large uncertainties at dense low-temperature state points. This effect is particularly important when highly dipolar fluids with large elongations are considered. As the uncertainty of $\mu^{\mbox{\scriptsize res*}}$ directly influences the uncertainty of the vapour pressure, Widom's method would spoil any accurate calculation of vapour pressures at these state points. More precise values for the residual chemical potential can be obtained from the gradual insertion method \cite{shevkunov8824,nezbeda9139,vrabec0243}. This method was used here in order to increase the accuracy of vapour pressure calculations at lowest temperatures $0.55 \cdot T^*_{\rm c}$ of the systems with $L^* \in [0;1.0]$ and $\mu^{*2} \in [0;12]$, except for $L^*=0$ with $\mu^{*2}=12$. For the systems $L^*=0$ with $\mu^{*2}=12$, $L^*=0$ or $L^*=0.2$ with $\mu^{*2}=16$, and $L^*=0$ with $\mu^{*2}=20$ Widom's method was applied with an increased number of test particles and extended simulation run lengths at state points $0.55 \cdot T^*_{\rm c}$. It must be pointed out, that due to these difficulties at low temperature state points, VLE simulations of dipolar fluids are performed typically down to about $0.70 \cdot T^*_{\rm c}$. Below, special techniques are required for good results.

Configuration space sampling was done by $N\!pT$-molecular dynamics simulations with $N=864$ particles for both liquid and vapour simulations. The dimensionless integration time step was set to $\Delta t^*=0.0015$. Starting from a face-centred lattice arrangement every simulation run was given $10,000$ integration time steps to equilibrate. Data production was performed over $n=100,000$ ($n=200,000$ for the three systems mentioned above) integration time steps. At each production time step $2N$ ($10N$ for the three systems mentioned above) test particles were inserted in the liquid phase, and $N$ test particles were inserted in the vapour phase, in order to calculate the chemical potentials. The dimensionless dynamical parameter of $N\!pT$-MD-simulations ascribed to the box membrane according to Andersen's algorithm \cite{andersen8023} was set to $2 \cdot 10^{-4}$ for liquid simulations and to $10^{-6}$ for vapour simulations. The high value of $N$ allowed simulations in the vicinity of the critical point, and the high value of $n$ was used in order to obtain small statistical uncertainties.

The gradual insertion method for the calculation of residual chemical potentials $\mu^{\mbox{\scriptsize res*}}$ is an expanded ensemble method based on the Monte Carlo technique. The gradual insertion method as proposed in \cite{nezbeda9139} extended to $N\!pT$ ensembles \cite{vrabec0243} was used in this work. In comparison to Widom's test particle method, where real particles are inserted in the fluid, in gradual insertion one fluctuating particle is introduced, that undergoes changes in a predefined set of discrete states of coupling with all other real particles of the fluid. Preferential sampling is done in the vicinity of the fluctuating particle. This concept leads to considerably improved accuracy of the residual chemical potential. Gradual insertion simulations were performed with $N=864$ particles in the liquid phase at $0.55 \cdot T^*_{\rm c}$. Starting from a face-centred lattice arrangement every simulation run was given $5,000$ Monte Carlo loops to equilibrate. Data production was performed over $n=100,000$ Monte Carlo loops. One Monte Carlo loop is defined here as $N$ trial translations, $\left(2/3\right)N$ trial rotations, and one trial volume change. Further simulation parameters for runs with gradual insertion were taken from Vrabec et al. \cite{vrabec0243}.

For the 2CLJD fluid a hybrid equation of state (EOS) for the Helmholtz energy $F_{\rm 2CLJD}$ is available \cite{kriebel9815,saager9267,mecke9768},
 which is constructed as sum of the Helmholtz energy $F_{\rm 2CLJ}$ for the Lennard-Jones part of the potential model and the Helmholtz energy $F_{\rm D}$ for the dipolar contribution. In planning the present investigation this 2CLJD EOS was used to estimate the critical temperatures $T^*_{\rm c} \left( \mu^{*2},L^* \right)$ for the systems with $L^* \in [0;0.8]$ and $\mu^{*2} \in [0;12]$.

For all systems VLE data were calculated for temperatures from about 55 \% to about 95 \% of the critical temperature. Liquid simulations were performed for the whole temperature range, whereas vapour simulations were only performed at temperatures above $0.80 \cdot T^*_{\rm c,estd}$. Below this temperature the second virial coefficient is sufficient for the VLE calculations. Liquid simulations took about eleven hours, vapour simulations about three hours CPU time on a modern workstation (e.g., Compaq AlphaStation XP1000). Liquid simulations with gradual insertion took an average of about 200 hours CPU time.
\clearpage

\section{Simulation results of VLE data and critical data}

Table \ref{t2cljdvle} reports an extract of the VLE data obtained in this work. All 38 fluids are covered, but not all state points for which simulations were carried out are included in Table \ref{t2cljdvle} for the sake of brevity. For each state point, the vapour pressure $p_{\sigma}^*$, the saturated liquid density $\rho'^*$, the saturated vapour density $\rho''^*$, the residual saturated liquid enthalpy $h'^{\mbox{\scriptsize res*}}$, and the residual saturated vapour enthalpy $h''^{\mbox{\scriptsize res*}}$ are reported. The temperatures are about $T^* \approx 0.55 \cdot T^*_{\rm c}$, $0.80 \cdot T^*_{\rm c}$, and $0.95 \cdot T^*_{\rm c}$. Statistical uncertainties were determined with the method of Fincham et al. \cite{fincham8645} and the error propagation law. The full data is available at http://www.itt.uni-stuttgart.de/molsim.html.

\begin{figure}[ht]
\epsfig{file=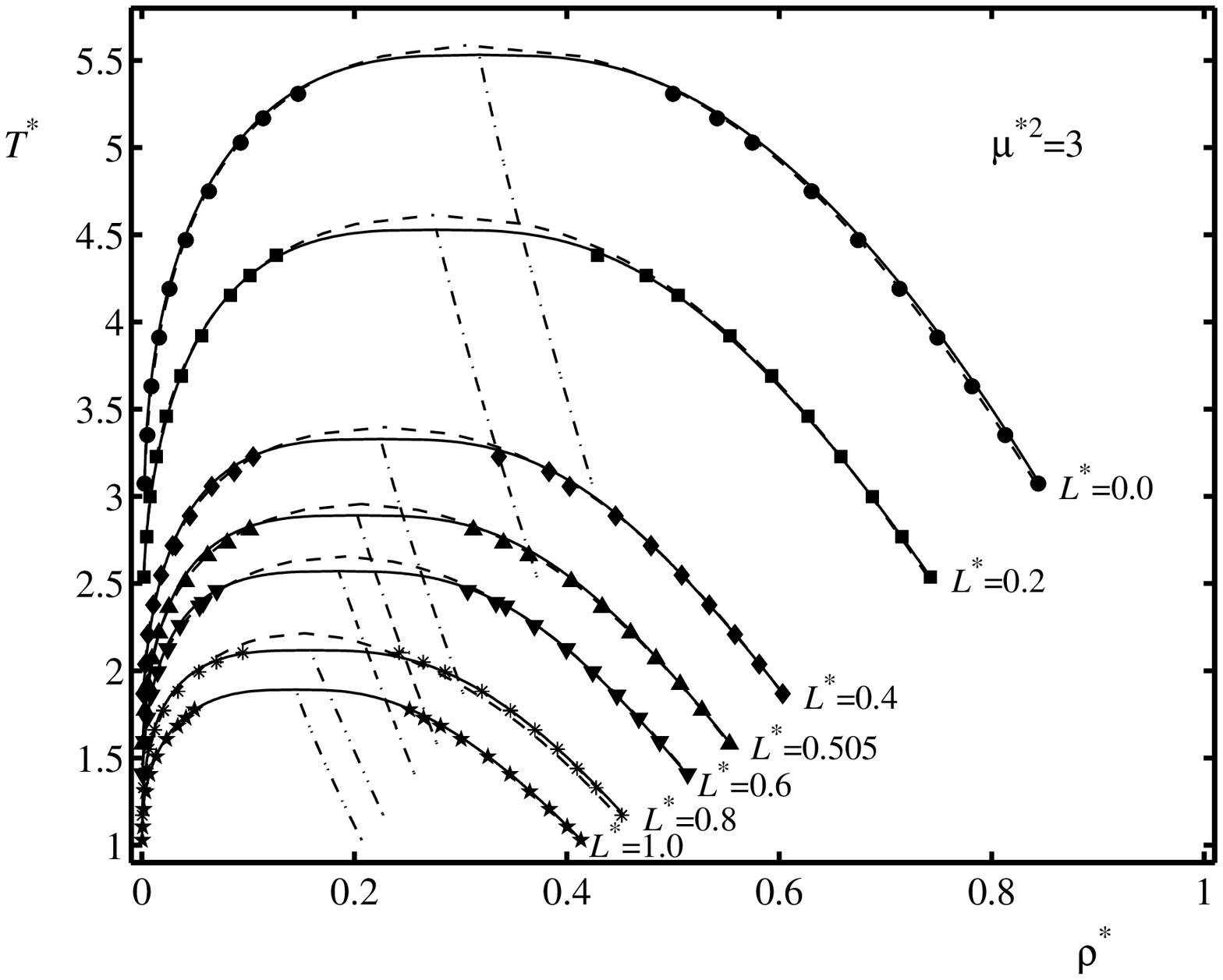,scale=1,angle=90}
\caption[Temperature--density coexistence curves of the 2CLJD model fluids with $\mu^{*2}=3$. Symbols: simulation data. {\mbox{\bf -----} cor\-re\-la\-tion}, this work. {\mbox{\bf -- $\cdot$ -- $\cdot$} rec\-ti\-li\-ne\-ar} diameter from correlation, this work. {\mbox{\bf -- -- --} 2CLJD EOS} \cite{kriebel9815,saager9267,mecke9768}, not valid beyond $L^*>0.8$. All error bars are within symbol size.]{}
\label{xa2Da}
\end{figure}

\begin{figure}[ht]
\epsfig{file=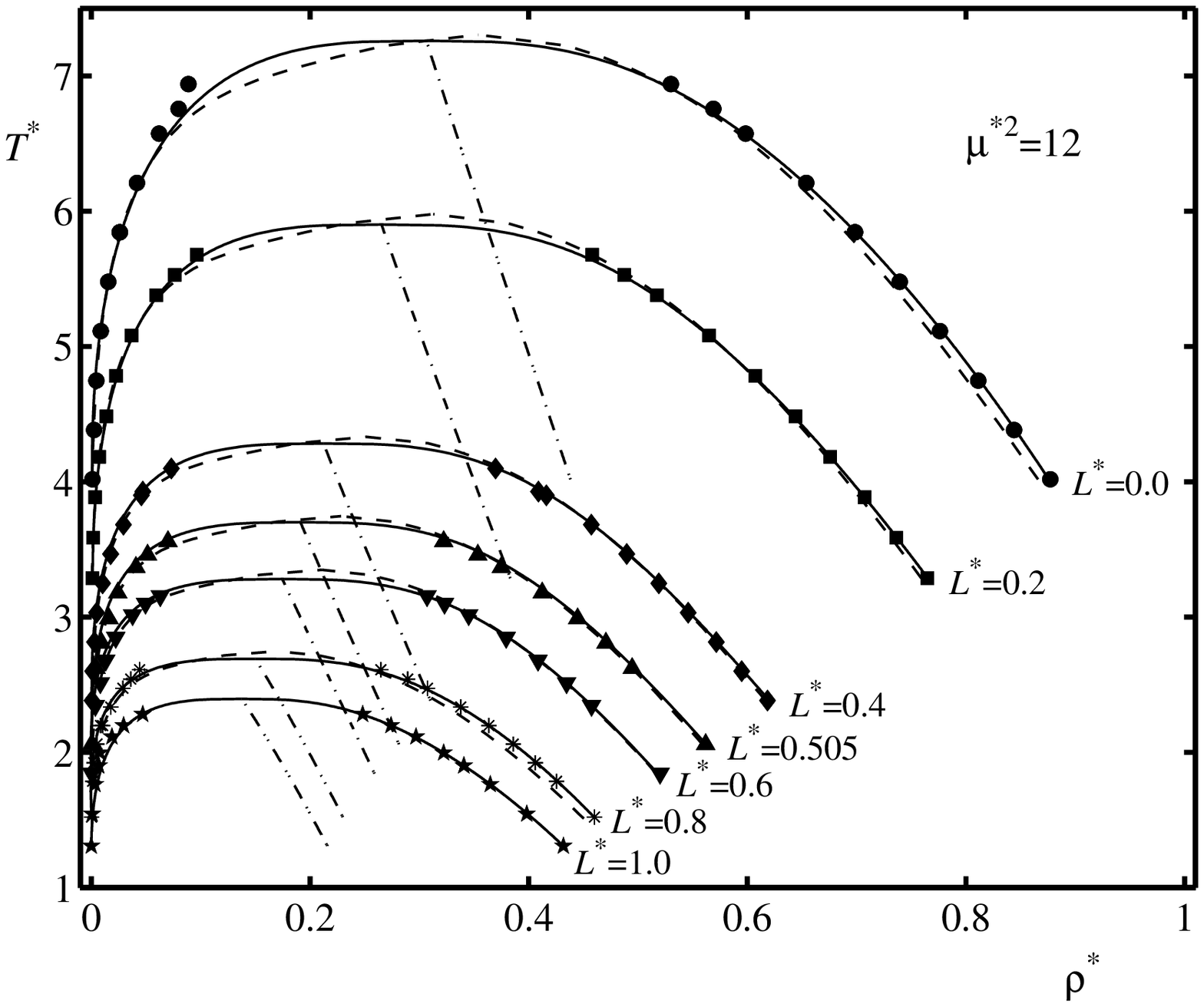,scale=1,angle=90}
\caption[Temperature--density coexistence curves of the 2CLJD model fluids with $\mu^{*2}=12$. Symbols: simulation data. {\mbox{\bf -----} cor\-re\-la\-tion}, this work. {\mbox{\bf -- $\cdot$ -- $\cdot$} rec\-ti\-li\-ne\-ar} diameter from correlation, this work. {\mbox{\bf -- -- --} 2CLJD EOS} \cite{kriebel9815,saager9267,mecke9768}, not valid beyond $L^*>0.8$. All error bars are within symbol size.]{}
\label{xa2Dd}
\end{figure}

\begin{figure}[ht]
\epsfig{file=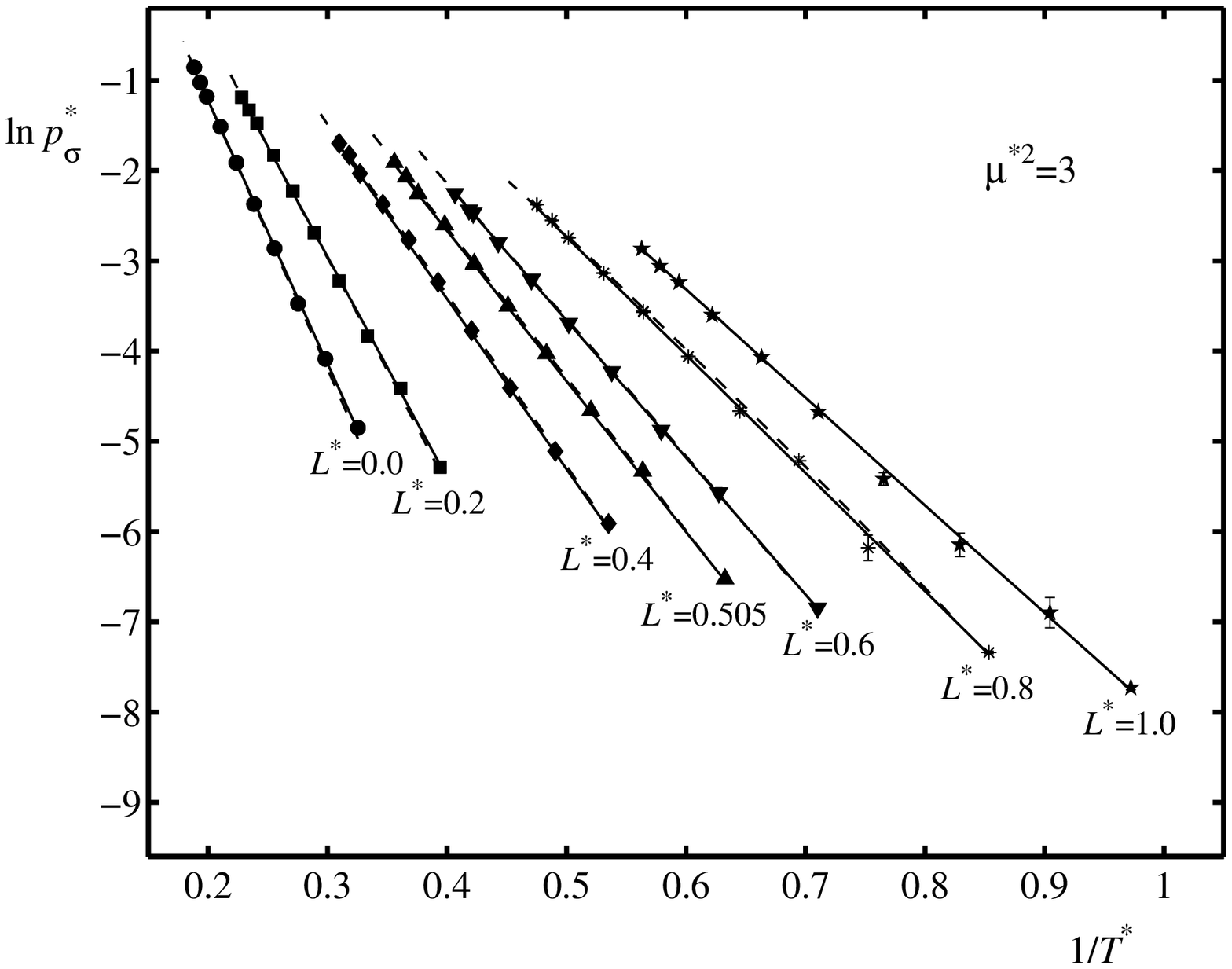,scale=1,angle=90}
\caption[Vapour pressure curves of the 2CLJD model fluids with $\mu^{*2}=3$. Symbols: simulation data. {\mbox{\bf -----} cor\-re\-la\-tion}, this work. {\mbox{\bf -- -- --} 2CLJD EOS} \cite{kriebel9815,saager9267,mecke9768}, not valid beyond $L^*>0.8$. Error bars are within symbol size, if not shown.]{}
\label{xa1Da}
\end{figure}

\begin{figure}[ht]
\epsfig{file=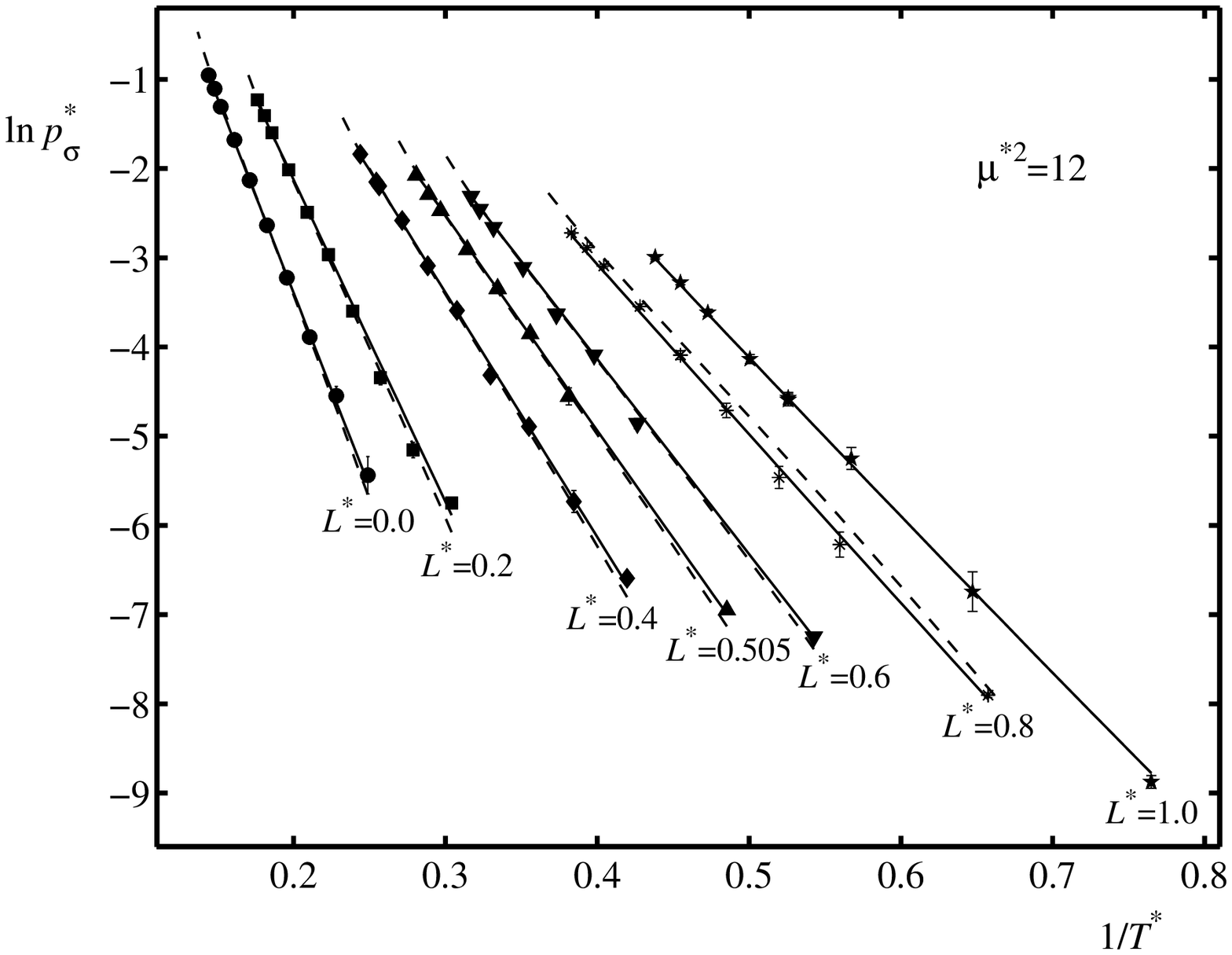,scale=1,angle=90}
\caption[Vapour pressure curves of the 2CLJD model fluids with $\mu^{*2}=12$. Symbols: simulation data. {\mbox{\bf -----} cor\-re\-la\-tion}, this work. {\mbox{\bf -- -- --} 2CLJD EOS} \cite{kriebel9815,saager9267,mecke9768}, not valid beyond $L^*>0.8$. Error bars are within symbol size, if not shown.]{}
\label{xa1Dd}
\end{figure}

Figs. \ref{xa2Da} -- \ref{xa1Dd} illustrate for $\mu^{*2}=3$ and $\mu^{*2}=12$ the strong influence of both the elongation and the dipolar momentum on the 2CLJD VLE data. Increasing the elongation or increasing the dipolar momentum strongly influences the shape of the density coexistence curve and the slope of the vapour pressure curve. At low temperatures, the high uncertainties of the vapour pressures of the systems with large elongations and strong dipolar momentum, cf. Fig. \ref{xa1Dd}, are due to the uncertain values of the chemical potentials obtained by Widom's test particle insertion in the liquid phases. Figs. \ref{xa1Da} and \ref{xa1Dd} illustrate, at lowest temperatures, the considerable decrease of the uncertainties of the vapour pressure when the chemical potential in the liquid phase is obtained with much lower uncertainties by gradual insertion.

The comparison of the 2CLJD VLE data from the present work to the 2CLJD EOS reveals systematic deviations. This can best be seen in Figs. \ref{xa2Dd} and \ref{xa1Dd}. For models with small dipolar momenta and for models with elongations near $L^*=0.505$ the 2CLJD VLE data from the simulations and that from the EOS agree well. This is due to the fact, that the dipolar contribution to the EOS is based on data of a 2CLJD model fluid with \mbox{$L^*=0.505$}. However, for other state points the EOS sometimes deviates considerably from the simulation data. In most cases the EOS underestimates the saturated liquid densities and overestimates the saturated vapour densities. The vapor pressures from the 2CLJD EOS generally agree well with simulation data from this work, except for model fluids with high dipolar momenta and large elongations, where the EOS overestimates the vapour pressures (cf. Fig. \ref{xa1Dd}, the EOS is not valid beyond $L^*=0.8$).

Critical data were determined here with the method of Lotfi et al. \cite{lotfi9213}, who, by simple means, found reliable critical data for the 1CLJ model fluid. It is known that the density--temperature dependence near the critical point is well described by $\rho^* \sim \left( T^*_{\rm c}-T^*\right)^{1/3}$, as given by Guggenheim \cite{guggenheim4525,rowlinson1969}. Eqs. (\ref{Trho1corrcore}) and (\ref{Trho2corrcore}), suggested by Lotfi et al. \cite{lotfi9213}, were used for the correlation of the saturated densities from simulation
\begin{eqnarray}
\rho'^*=\rho^*_{\rm c} + C_1 \cdot (T^*_{\rm c}-T^* )^{1/3} + C'_2 \cdot (T^*_{\rm c}-T^*) + C'_3 \cdot (T^*_{\rm c}-T^*)^{3/2},
\label{Trho1corrcore} \\
\rho''^*=\rho^*_{\rm c} - C_1 \cdot (T^*_{\rm c}-T^* )^{1/3} + C''_2 \cdot (T^*_{\rm c}-T^*) + C''_3 \cdot (T^*_{\rm c}-T^*)^{3/2}.
\label{Trho2corrcore}
\end{eqnarray}
The simultaneous fit of saturated liquid and saturated vapour densities yields not only the coefficients $C_1$, $C'_2$, $C'_3$, $C''_2$, $C''_3$, but also the critical data $\rho^*_{\rm c}$, $T^*_{\rm c}$. The critical temperatures and densities for the 2CLJD model fluids are listed in Table \ref{2CLJDcdomtab}. In most cases critical temperatures obtained from the simulation data are lower than those estimated by the 2CLJD EOS. Table \ref{2CLJDcdomtab} also contains the critical compressibility factor $Z_{\rm c}=p^*_{\rm c}/\left(\rho^*_{\rm c} T^*_{\rm c}\right)$, which is a non-reduced property of the 2CLJD fluids. It is therefore of particular interest for comparisons to real dipolar fluids.

The uncertainties of $T^*_{\rm c}$ and $\rho^*_{\rm c}$ are similar as those found in an analogous investigation of the two-centre Lennard-Jones plus pointquadrupole fluid \cite{stoll0133}. They are estimated to be $\sigma\!\left( T^*_{\rm c} \right) \approx 0.005$ and $\sigma\!\left( \rho^*_{\rm c} \right) \approx 0.0005$, and it is concluded, that the critical temperatures calculated by this method are certain up to the second, the critical densities up to the third digit after the decimal point.
\clearpage

\section{Global correlation of VLE data}

In order to obtain VLE data for the whole range of $\mu^{*2}$, $L^*$ and $T^*$ the molecular simulation data from this work were globally correlated. The critical data \mbox{$T^*_{\rm c}(\mu^{*2},L^*)$}, \mbox{$\rho^*_{\rm c}(\mu^{*2},L^*)$}, the saturated liquid density \mbox{$\rho'^*(\mu^{*2},L^*,T^*)$} and the vapour pressure
\linebreak
\mbox{$p^*_{\sigma}(\mu^{*2},L^*,T^*)$} are the key VLE data for an adjustment to real fluids. The adequate shape of the temperature--density coexistence curve was achieved by simultaneously correlating the functions \mbox{$\rho'^*\left(\mu^{*2},L^*,T^*\right)$} and \mbox{$\rho''^*\left(\mu^{*2},L^*,T^*\right)$}. It was not in the scope of the present investigation to construct a new 2CLJD EOS. The correlation developed here is not designed to compete with an EOS. It shall merely be a working tool for a restricted field of application, namely the adjustment of model parameters to data of real fluids as it has been suggested by Vrabec et al. \cite{vrabec0112}. The vapour pressure correlation is also used to verify the thermodynamic consistency of the VLE data from simulations by the means of the Clausius-Clapeyron equation. Moreover, the correlations are useful for comparisons with results of other investigators. Details of the correlation method are described in the Appendix B which also contains the resulting correlation functions.

\subsection{Critical properties}

\begin{figure}[ht]
\epsfig{file=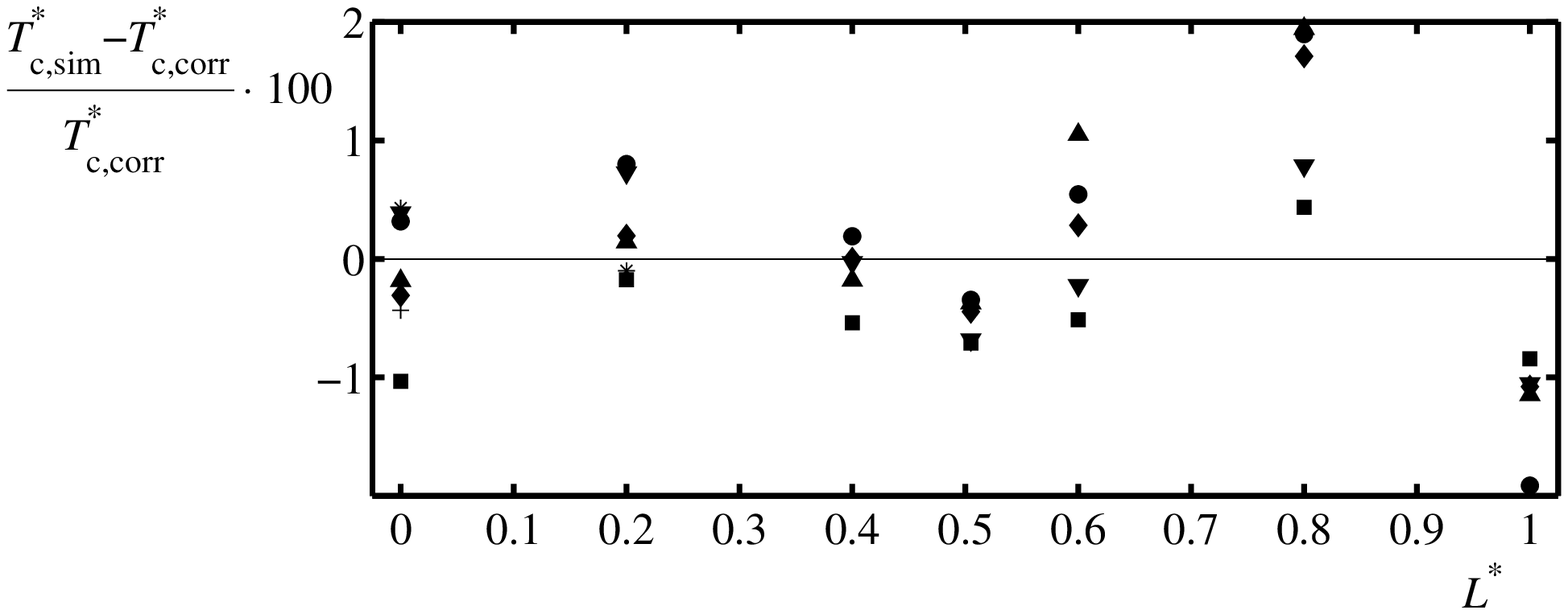,scale=1,angle=90}
\epsfig{file=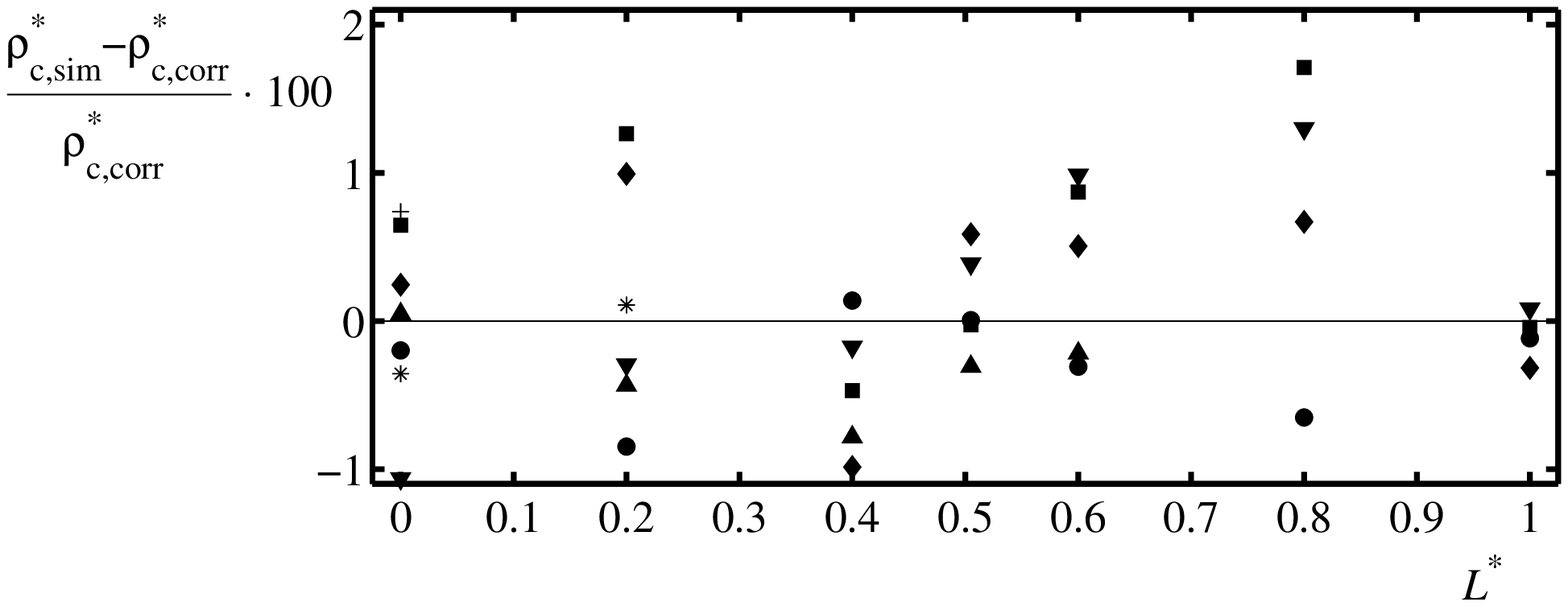,scale=1,angle=90}
\caption[Relative deviations of simulation data to the correlation. Top: critical temperatures $(T^*_{\rm c,sim}-T^*_{\rm c,corr})/T^*_{\rm c,corr}$. Bottom: critical densities $(\rho^*_{\rm c,sim}-\rho^*_{\rm c,corr})/\rho^*_{\rm c,corr}$. Symbols are: $\mu^{*2}=0$ ({\large $\bullet$}), $3$ ({\Huge $\centerdot$}), $6$ ({\footnotesize $\blacklozenge$}), $9$ ($\blacktriangle$), $12$ ($\blacktriangledown$), $16$ ($*$), $20$ ($+$).]{}
\label{xrTcRhoc}
\end{figure}

The correlation functions $T^*_{\rm c}(\mu^{*2},L^*)$ and $\rho^*_{\rm c}(\mu^{*2},L^*)$ were assumed to be linear combinations of elementary functions, cf. Eq. (\ref{lincomb}) in the Appendix B. The elementary functions and their coefficients are given in Table \ref{corrtab} in the Appendix B. The quality of the correlations can be studied in Fig. \ref{xrTcRhoc}. Most relative deviations of the critical temperatures are within 1 \%. The critical densities are represented with roughly the same quality. It should be mentioned that possible systematic errors may be introduced to the critical data by the choice of the exponent $1/3$ in the second term of Eqs. (\ref{Trho1corrcore}) and (\ref{Trho2corrcore}).

\subsection{Saturated densities, vapour pressure}

The temperature--saturated density correlations are based on Eqs. (\ref{Trho1corrcore}) and (\ref{Trho2corrcore}), which have five adjustable parameters $C_1$, $C'_2$, $C'_3$, $C''_2$, and $C''_3$. Another three parameters are introduced by the correlation of the vapour pressure, cf. Eq. (\ref{pTcorr}) in the Appendix B. Again the functions describing the dependency of the correlation parameters on $\mu^{*2}$ and $L^*$ were assumed to be linear combinations of elementary functions. The elementary functions and their coefficients are given in Table \ref{corrtab} in the Appendix B.

\begin{figure}[ht]
\epsfig{file=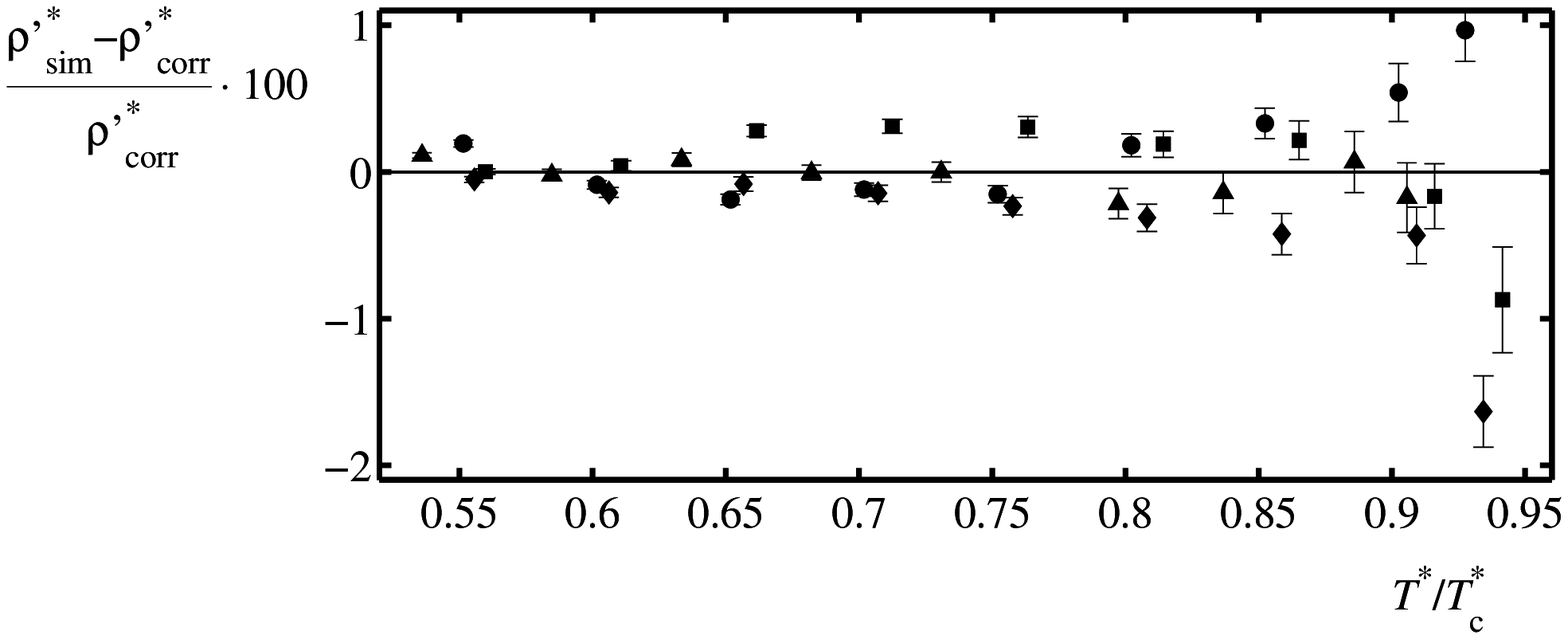,scale=1,angle=90}
\epsfig{file=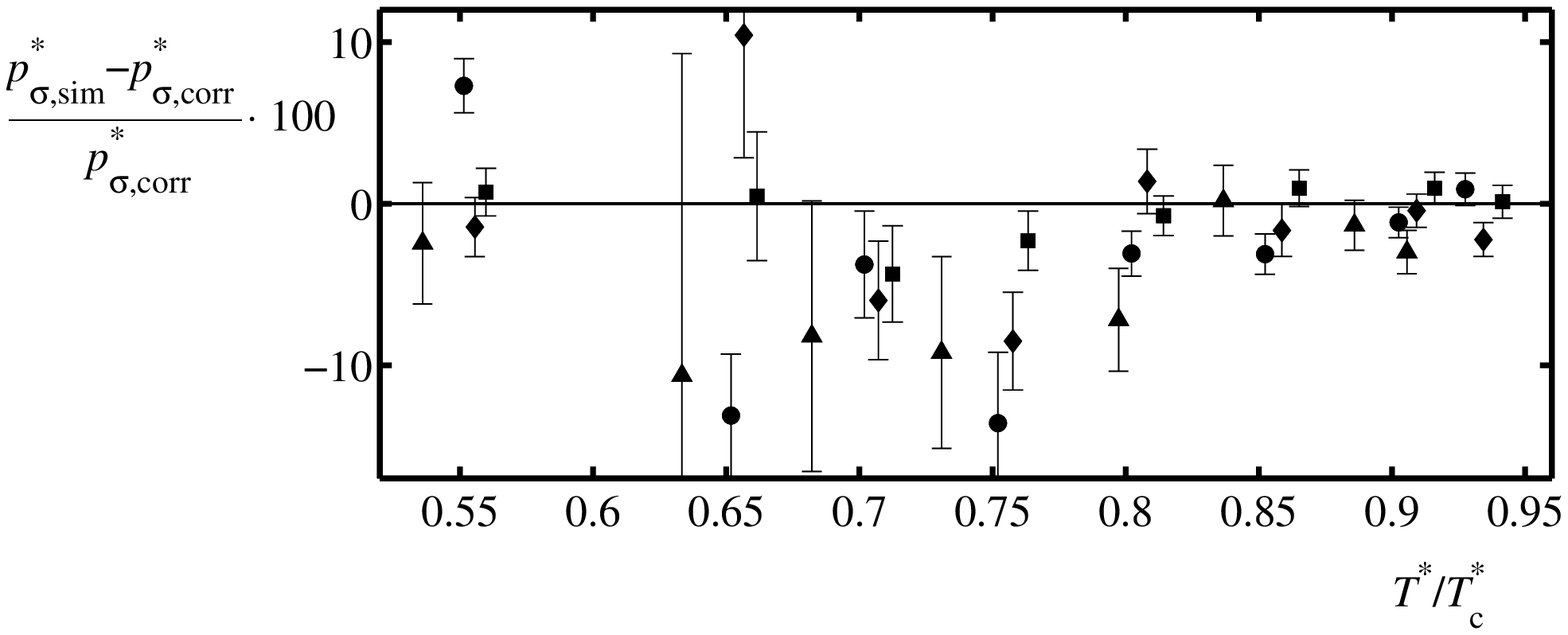,scale=1,angle=90}
\caption[Relative deviations of simulation data to the correlation. Top: saturated liquid densities $(\rho'^*_{\rm sim}-\rho'^*_{\rm corr})/\rho'^*_{\rm corr}$. Bottom: vapour pressures $(p^*_{\sigma \rm,sim}-p^*_{\sigma,\rm corr})/p^*_{\sigma,\rm corr}$. Symbols are: Stockmayer fluid (1CLJD) with $\mu^{*2}=9$ ({\large $\bullet$}), 2CLJD with $\mu^{*2}=6$ and $L^*=0.4$ ({\huge $\centerdot$}), 2CLJD with $\mu^{*2}=9$ and $L^*=0.505$ ({\footnotesize $\blacklozenge$}), and 2CLJD with $\mu^{*2}=6$ and $L^*=1$ ($\blacktriangle$).]{}
\label{xrRlpsDLT}
\end{figure}

A comparison between the correlations and the simulation data can be seen in Figs. \ref{xa2Da} -- \ref{xa1Dd}. A more detailed comparison is given in Fig. \ref{xrRlpsDLT}. For some of the 2CLJD fluids investigated here, the relative deviations of the simulation results from the correlations are shown. Typically, the saturated liquid density correlation has the largest deviations for the highest temperature of each model, which is due to the large uncertainties in the near critical region. The correlation shows relative deviations in the range of 0.4 \% for mid temperatures which are most important for adjustments to real fluids.

Relative deviations of the saturated vapour densities are not illustrated here. The vapour density correlation should not be used below $0.60 \cdot T_{\rm c}(\mu^{*2},L^*)$. At low temperatures the correlation is not useful as it does not capture the limiting case of the ideal gas, which is independent of the parameters $\mu^{*2}$ and $L^*$.

In most cases, the vapour pressure correlation represents the simulation data within their uncertainties. It has to be mentioned, that, except for simulations with gradual insertion, simulations at low temperature state points yield vapour pressures with increased uncertainties due to the uncertain values of the chemical potential obtained by Widom's test particle insertion in dense liquid phases.

By extrapolating the vapour pressure correlation slightly to the critical point, the acentric factor \cite{pitzer5534}
\begin{eqnarray}
\omega \left( \mu^{*2},L^* \right)=-{\rm log}_{10}\frac{p^* \left( \mu^{*2},L^*,0.7 \cdot T^*_{\rm c}\right)}{p^*_{\rm c} \left( \mu^{*2},L^* \right)}-1
\label{omega}
\end{eqnarray}
can be calculated from the correlations discussed before. Table \ref{2CLJDcdomtab} contains the critical pressures and the acentric factors for the 2CLJD model fluids calculated on the basis of these correlations.
\clearpage

\section{Discussion}

\subsection{Comparison to results of other authors}

\begin{figure}[ht]
\epsfig{file=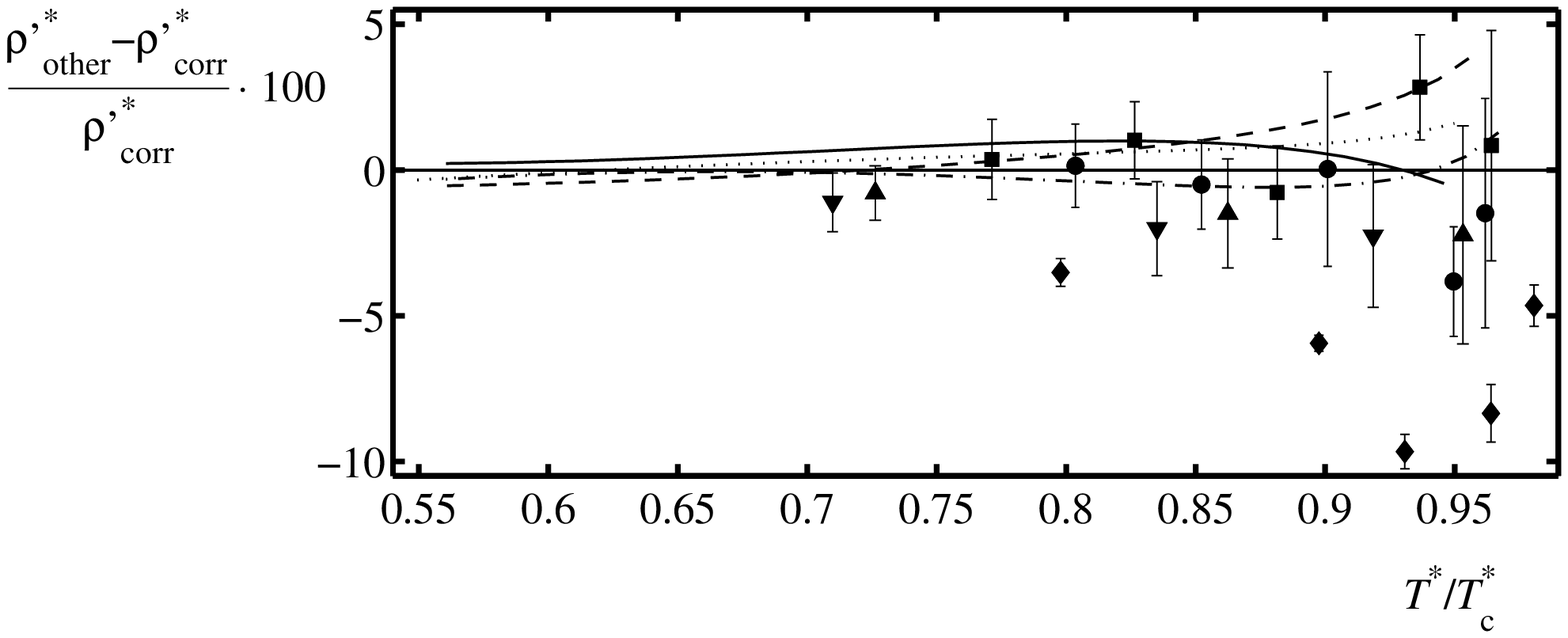,scale=1,angle=90}
\epsfig{file=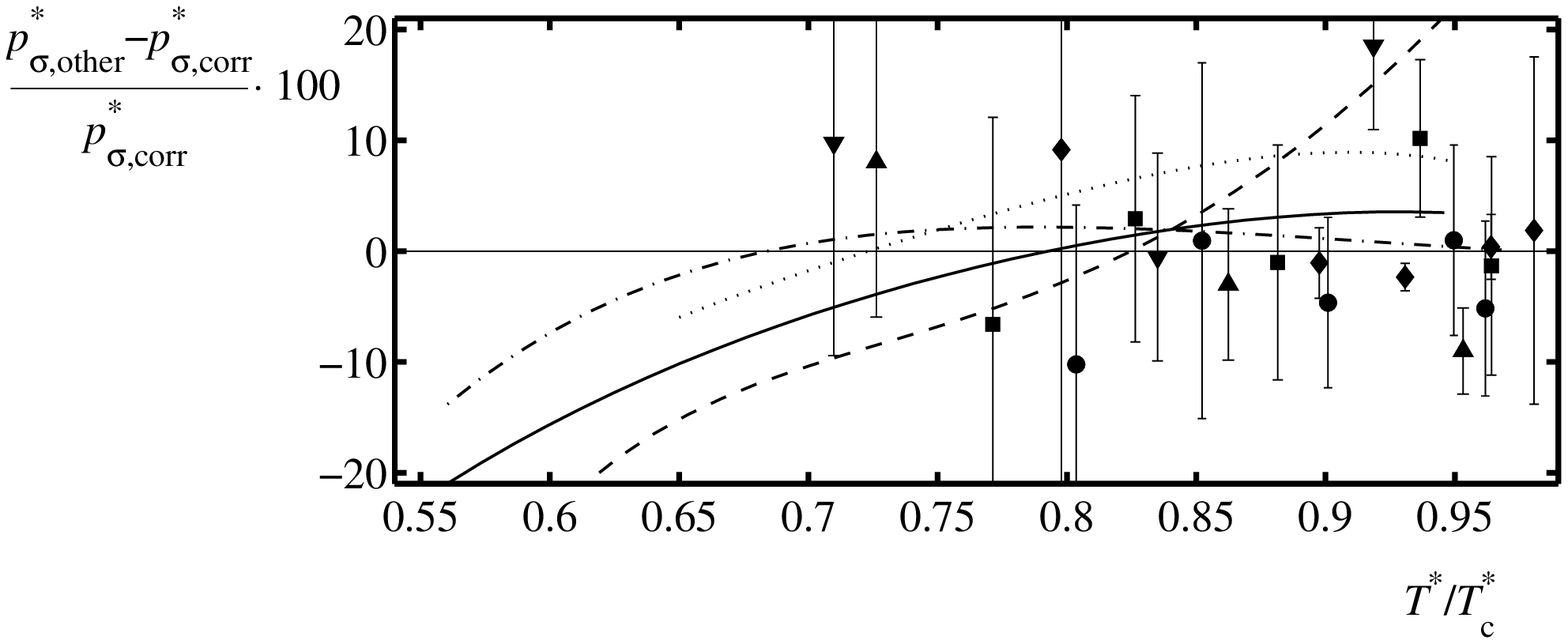,scale=1,angle=90}
\caption[Relative deviations between simulation data of other authors and correlations based on the simulations from the present work. Top: saturated liquid densities $\rho'^*=(\rho'^*_{\rm other}-\rho'^*_{\rm corr})/\rho'^*_{\rm corr}$. Bottom: vapour pressures $(p^*_{\sigma, \rm other}-p^*_{\sigma, \rm corr})/p^*_{\sigma, \rm corr}$. Simulation data from other authors: \cite{leeuwen9327} Stockmayer fluid (1CLJD) with $\mu^{*2}=16$ ({\large $\bullet$}); \cite{leeuwen9327} Stockmayer fluid (1CLJD) with $\mu^{*2}=12$ ({\huge $\centerdot$}); \cite{lisal9916} 2CLJD with $\mu^{*2}=6$ and $L^*=0.505$ \mbox{({\bf -- $\cdot$ --})}; \cite{lisal9719} 2CLJD with $\mu^{*2}=12$ and $L^*=0.67$ ({\bf -- -- --}); \cite{lisal9719} 2CLJD with $\mu^{*2}=12$ and $L^*=0.3292$ ({\bf $\cdot \cdot \cdot \cdot \cdot$}); \cite{lisal9719} 2CLJD with $\mu^{*2}=12$ and $L^*=0.22$ ({\bf ------}); \cite{dubey9399} 2CLJD with $\mu^{*2}=9$ and $L^*=0.6$ ($+$); \cite{lisal0036} 2CLJD with $\mu^{*2}=9$ and $L^*=1$ ($\blacktriangle$); \cite{lisal0036} 2CLJD with $\mu^{*2}=12$ and $L^*=1$ ($\blacktriangledown$). Error bars show the uncertainties, if they have been indicated.]{}
\label{xvergl_1_3}
\end{figure}

The results from the present study are compared here to the simulation results of other authors. Fig. \ref{xvergl_1_3} presents results for the relative deviations for the saturated liquid density $(\rho'^*_{\rm other}-\rho'^*_{\rm corr})/\rho'^*_{\rm corr}$ and the vapour pressure $(p^*_{\sigma,\rm other}-p^*_{\sigma,\rm corr})/p^*_{\sigma,\rm corr}$. The saturated liquid densities of other authors agree in almost all cases within the combined uncertainties of our correlation and of the simulation results, cf. Fig. \ref{xvergl_1_3}, top. Near critical points of course show larger deviations. Molecular simulations with considerably lower particle numbers yield slightly lower densities, thus causing systematic negative deviations, as can be observed, for example, for data from L\'{\i}sal et al. \cite{lisal0036}. Also for the vapour pressure good agreement is observed, as large uncertainties of the literature data have to be assumed if they have not been specified, cf. Fig. \ref{xvergl_1_3}, bottom.

Present 2CLJD VLE data generally show much lower statistical uncertainties than those of other authors, and they have the advantage that uniformly the same simulation method was used to produce them over a large parameter range.

\subsection{Thermodynamic consistency test}

The thermodynamic consistency of the simulation data was checked with the Clausius-Clapeyron equation
\begin{eqnarray}
\frac{\partial {\rm ln}p^*_{\sigma}}{\partial T^*}=\frac{\Delta h^*_{\rm v}}{p^*_{\sigma} T^* \left( 1/\rho''^*-1/\rho'^* \right)}.
\label{clcl}
\end{eqnarray}
The vapour pressure correlation from this work was used to evaluate the left hand side of Eq. (\ref{clcl}). The right hand side of Eq. (\ref{clcl}) was calculated from the simulation data of this work, the uncertainty was calculated by the error propagation law. The requirements of Eq. (\ref{clcl}) are fulfilled for almost all temperatures within the uncertainties of the right hand side of Eq. (\ref{clcl}). Hence, it is concluded that the data from this work are thermodynamically consistent.

\subsection{Locus of the critical point}

The influence of the dipole $\mu^{*2}$ and the elongation $L^*$ on the critical properties can be studied in Table \ref{2CLJDcdomtab} and in Figs. \ref{xa2Da} and \ref{xa2Dd}. The critical temperature $T^*_{\rm c}$ and the critical density $\rho^*_{\rm c}$ decrease with increasing $L^*$ at fixed dipolar momentum $\mu^{*2}$. They increase together with $\mu^{*2}$ at fixed elongation $L^*$. The absolute increase  $\Delta T^*_{\rm c} = T^*_{\rm c} \left( \mu^{*2},L^* \right) - T^*_{\rm c} \left( \mu^{*2}=0,L^* \right)$ of the critical temperature with increasing $\mu^{*2}$ is more important for molecules with small $L^*$ than for those with large $L^*$. In contrast, the relative increase $\Delta T^*_{\rm c}/T^*_{\rm c} \left( \mu^{*2}=0,L^* \right)$ of the critical temperature with increasing $\mu^{*2}$ only weakly depends on $L^*$. A comparison of this finding to results from Garz\'on et al. \cite{garzon9572}, who investigated VLE and critical data of Kihara fluids with axial pointdipole $\mu^*$ and rod length $L^*$, yields, that the thermodynamics of 2CLJD fluids and Kihara fluids with axial dipole are essentially different. Garz\'on et al. \cite{garzon9572} found, that the relative increase of the critical temperature of Kihara fluids with axial dipole distinctly depends on the rod length $L^*$ of the Kihara potential.

The critical pressure $p^*_{\rm c}$ of the 2CLJD fluid decreases strongly with increasing elongation $L^*$ for a fixed dipolar momentum $\mu^{*2}$. For a fixed elongation $L^*$, the critical pressure typically shows a maximum with increasing dipolar momentum $\mu^{*2}$, cf. Table \ref{2CLJDcdomtab}. This behaviour agrees with results from L\'{\i}sal et al. \cite{lisal9719,lisal9916} for the 2CLJD fluid.

As shown in Figs. \ref{xa1Da} and \ref{xa1Dd}, the absolute value of the slope of the function $\ln p^*_{\sigma}$ vs. $1/T^*$, i.e. the enthalpy of vaporization $\Delta h^*_{\rm v}$, increases with increasing $\mu^{*2}$ for a fixed elongation $L^*$. It decreases with increasing elongation $L^*$ for a fixed dipolar momentum $\mu^{*2}$. It can also be seen that the vapour pressure $p^*_{\sigma}$ of a model fluid of given elongation $L^*$ decreases when the dipolar momentum $\mu^{*2}$ is increased.

\subsection{Deviation from principle of corresponding states}

\begin{figure}[ht]
\epsfig{file=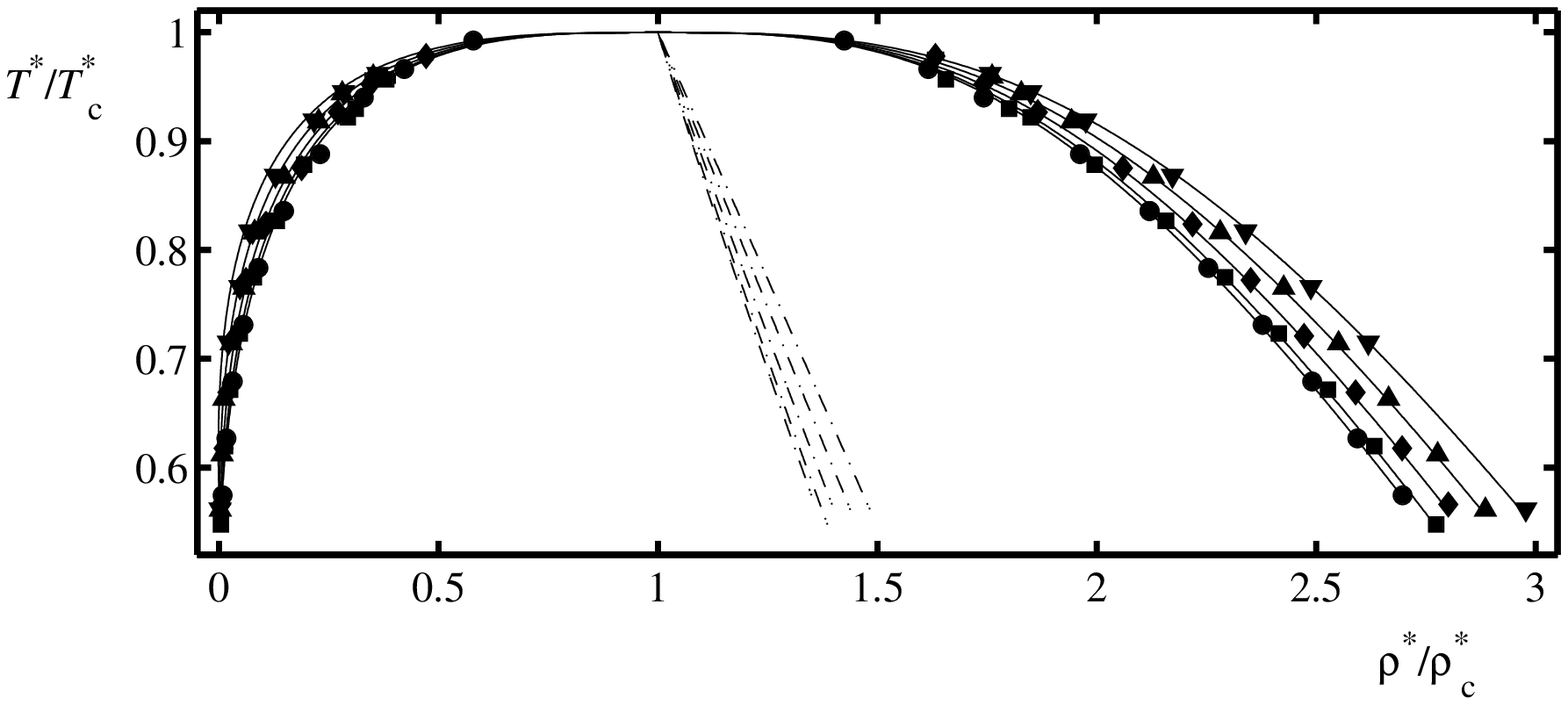,scale=1,angle=90}
\epsfig{file=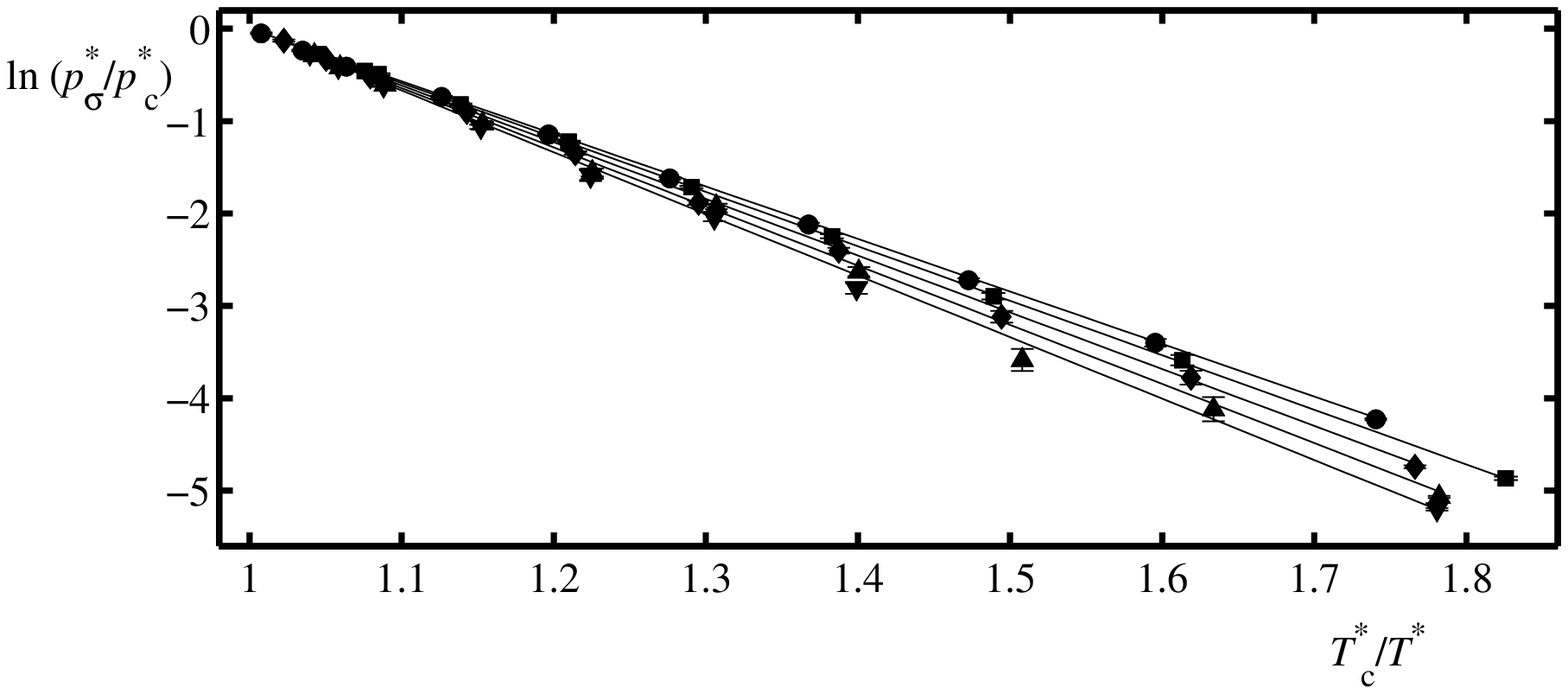,scale=1,angle=90}
\caption[Deviation from the principle of corresponding states caused by the influence of the dipole. Top: saturated densities. Bottom: vapour pressures. Elongation is $L^*=0.6$, with $\mu^{*2}=0$ ({\large $\bullet$}), $3$ ({\Huge $\centerdot$}), $6$ ({\footnotesize $\blacklozenge$}), $9$ ($\blacktriangle$), $12$ ($\blacktriangledown$).]{}
\label{xa2a1L06_ks}
\end{figure}

Both the presence of a dipole and molecular anisotropy cause deviations from the simple principle of corresponding states \cite{garzon9572,lisal9719}, i.e. neither the plots $\ln \left(p^*_{\sigma}/p^*_{\rm c}\right)$ vs. $T^*_{\rm c}/T^*$ nor the plots $T^*/T^*_{\rm c}$ vs. $\rho^*/\rho^*_{\rm c}$ show unique curves regardless of the values of $\mu^{*2}$ and $L^*$. The widening effect of the dipole on the density coexistence curve is shown in top of Fig. \ref{xa2a1L06_ks} for various 2CLJD fluids with $L^*=0.6$, and the displacement of the vapour pressure curves due to the dipole is shown in bottom of Fig. \ref{xa2a1L06_ks}. These effects have already been described by Lupkowski and Monson \cite{lupkowski8953} and L\'{\i}sal et al. \cite{lisal9719} for dipolar two-centre Lennard-Jones fluids.

These deviations from the simple principle of corresponding states are reflected by the behaviour of the acentric factor $\omega$ vs. $\mu^{*2}$ and $L^*$, cf. Eq. (\ref{omega}) and Table \ref{2CLJDcdomtab}. The more a fluid deviates from that principle, the higher $\omega$ will be.

A principle of corresponding states for the relative increase of the critical temperature has been derived by Garz\'on et al. \cite{garzon9572,garzon9441} for dipolar and quadrupolar Kihara fluids. In analogy to results in a comprehensive investigation of the 2CLJQ fluid \cite{stoll0133}, where that principle of corresponding states could not be confirmed for the 2CLJQ fluid, in this work, that principle could not be confirmed for the 2CLJD fluid, either. This is a further clear hint on considerable differences between the thermodynamics of two-centre Lennard-Jones fluids and Kihara fluids.
\clearpage

\section{Conclusion}

The present paper aims at the qualitative and quantitative improvement of available VLE data of the 2CLJD model fluid. In a systematic investigation the two parameters $\mu^{*2}$ and $L^*$ of that model fluid were varied in the ranges $0 \le \mu^{*2} \le 20$ and $0 \le L^* \le 1.0$ respectively. A total of 38 model fluids were studied in detail, including the non-polar and spherical cases. The $N\!pT+$ Test Particle Method was applied for the production of VLE data in the temperature range of about $0.55 \cdot T^*_{\rm c} \le T^* \le 0.95 \cdot T^*_{\rm c}$ for all model fluids. At the lowest temperature of systems with $\mu^{*2} \le 12$ the chemical potentials were calculated by gradual insertion. The comparison of data from this work to a hybrid 2CLJD EOS \cite{kriebel9815,saager9267,mecke9768} reveals some shortcomings of that EOS. Critical data for all 38 systems were obtained from individual fits of the saturated densities with Guggenheim's equations.

In order to obtain useful tools for adjustments of model parameters of the 2CLJD fluid to experimental data of real fluids, global correlations of the critical data, the saturated density coexistence curve, and the vapour pressure curve of the 2CLJD fluid were developed. Data from this work agree well with results of other investigators. In most cases, however, data from this work have lower uncertainties, furthermore the whole data set is self-consistent.

Using the correlations from this work, the influence of $\mu^{*2}$ and $L^*$ on the locus of the critical point and on the shapes of the saturated density coexistence curves and of the vapour pressure curves were studied. Deviations from the principle of corresponding states are due to the presence of polarity and anisotropy.

The correlations resulting from this work will be used to develop molecular models for real dipolar fluids.
\clearpage

\section{List of Symbols}

\begin{tabular}{ll}
$a$ & interaction site counting index \\
$b$ & interaction site counting index \\
$C$ & coefficient of correlation function \\
$c$ & coefficient of correlation function \\
$c$ & constant in set of elementary functions \\
$F$ & Helmholtz energy \\
$F$ & function to minimize \\
$G$ & function to minimize \\
$h$ & enthalpy \\
$i$ & data point counting index \\
$i$ & elementary function counting index \\
$i$ & particle counting index \\
$j$ & elementary function counting index \\
$j$ & particle counting index \\
$k$ & Boltzmann constant \\
$k$ & elementary function counting index \\
$L$ & molecular elongation \\
$\ell$ & simplified notation for $L^*$ \\
$m$ & mass of particle \\
$m$ & simplified notation for $\mu^{*2}$ \\
$N$ & number of particles \\
$n$ & number of time steps \\
$n$ & number of Monte Carlo loops \\
$p$ & pressure \\
$r$ & site-site distance \\
$r_{\rm c}$ & centre-centre cut-off radius \\
$T$ & temperature \\
$t$ & time \\
$u$ & pair potential \\
$u$ & internal energy \\
$w$ & virial \\
$y$ & linear combination of elementary functions \\
$Z$ & compressibility factor \\
\end{tabular}
\clearpage

\noindent
\begin{tabular}{ll}
$\alpha$ & coefficient of elementary function \\
$\beta$ & coefficient of elementary function \\
$\gamma$ & coefficient of elementary function \\
$\gamma_{ij}$ & angle between two dipole vectors \\
$\Delta h_{\rm v}$ & enthalpy of vapourisation \\
$\Delta T_{\rm c}$ & absolute increase of the critical temperature \\
$\Delta t$ & integration time step\\
$\delta$ & statistical uncertainty \\
$\epsilon$ & Lennard-Jones energy parameter \\
$\epsilon_{\rm s}$ & relative permittivity of dielectric continuum \\
$\epsilon_0$ & permittivity of the vacuum \\
$\theta$ & angle of nutation \\
$\mu$ & dipolar momentum \\
$\mu$ & chemical potential \\
$\xi$ & elementary function \\
$\rho$ & density \\
$\sigma$ & Lennard-Jones size parameter \\
$\sigma$ & standard deviation \\
$\chi$ & elementary function \\
$\psi$ & elementary function \\
$\omega$ & acentric factor \\
\end{tabular}

\vspace{0.7cm}

\noindent
\textbf{\large Vector properties} \\[0.6cm]
\begin{tabular}{ll}
$\bm E$ & electric field vector \\
$\bm r$ & position vector \\
$\bm \mu$ & dipole vector \\
$\bm \tau$ & torque vector \\
$\bm \omega$ & orientation vector \\
\end{tabular}
\clearpage

\noindent
\textbf{\large Subscripts} \\[0.6cm]
\begin{tabular}{ll}
c & property at critical point \\
corr & from present correlation \\
D & dipole \\
estd & estimated value \\
other & from other authors \\
RF & reaction field \\
sim & from present simulation \\
$\sigma$ & vapour-liquid coexistence \\
2CLJ & two-centre Lennard-Jones \\
2CLJD & two-centre Lennard-Jones plus pointdipole \\
\end{tabular}

\vspace{0.7cm}

\noindent
\textbf{\large Superscripts} \\[0.6cm]
\begin{tabular}{ll}
*     & reduced \\
$'$   & on bubble line \\
$''$  & on dew line \\
id  & ideal gas \\
res & residual property \\
tot & for all particles \\
\end{tabular}
\clearpage

\section{Acknowledgments}
The authors thank Prof. J. Fischer, Vienna, for fruitful discussions. We gratefully acknowledge financial support by Deutsche Forschungsgemeinschaft, Son\-der\-for\-schungs\-be\-reich 412, University of Stuttgart.
\clearpage


\clearpage

\begin{table}[ht]
\noindent
\caption[]{Vapour-liquid equilibrium data. Extract from simulation results for 38 model fluids for low, mid, and high temperatures. At the low temperatures, the second virial coefficient was used for the vapour phase. At the temperatures marked by \dag, the data are based on chemical potentials obtained by gradual insertion. For the remaining temperatures, the data are based on chemical potentials obtained by Widom's method. The numbers in parentheses indicate the uncertainties of the last decimal digits.}
\label{t2cljdvle}
\medskip
\begin{center}
{\footnotesize
\begin{tabular}{l|ccccc}\hline\hline
{\normalsize ~$\!T^*$} &
{\normalsize $p_{\sigma}^*$} &
{\normalsize $\rho'^*$} & {\normalsize $\rho''^*$} &
{\normalsize $h'^{\mbox{\scriptsize res*}}$} &
{\normalsize $h''^{\mbox{\scriptsize res*}}$} \\ \hline
\multicolumn{6}{l}{{\normalsize $L^*=0$, $\mu^{*2}=0$}} \\ \hline
2.92600\dag  &   0.00883    (10) &   0.82998    (20) &   0.00310\0   (4) & -26.8419\0   (57) &  -0.2279\0   (27) \\
4.25600      &   0.1469\0   (14) &   0.66404    (55) &   0.04347    (53) & -22.457\0\0  (15) &  -2.459\0\0  (33) \\
5.05400      &   0.4120\0   (34) &   0.4925\0   (31) &   0.1475\0   (38) & -17.768\0\0  (77) &  -6.98\0\0\0 (17) \\ \hline
\multicolumn{6}{l}{{\normalsize $L^*=0$, $\mu^{*2}=3$}} \\ \hline
3.07290\dag  &   0.00783\0   (8) &   0.84391    (19) &   0.00262\0   (3) & -30.0653\0   (79) &  -0.2347\0   (26) \\
4.46960      &   0.1476\0   (16) &   0.67441    (56) &   0.04122    (46) & -24.795\0\0  (17) &  -2.621\0\0  (26) \\
5.30770      &   0.4245\0   (37) &   0.5001\0   (30) &   0.1472\0   (38) & -19.507\0\0  (77) &  -7.70\0\0\0 (18) \\ \hline
\multicolumn{6}{l}{{\normalsize $L^*=0$, $\mu^{*2}=6$}} \\ \hline
3.33250\dag   &   0.00599\0   (8) &   0.85910    (17) &   0.00184\0   (3) & -35.1985\0   (96) &  -0.2665\0   (37) \\
4.84720       &   0.1352\0   (14) &   0.68964    (48) &   0.03466    (45) & -29.143\0\0  (16) &  -2.970\0\0  (43) \\
5.75065       &   0.4053\0   (35) &   0.5200\0   (20) &   0.1224\0   (31) & -23.346\0\0  (59) &  -8.11\0\0\0 (20) \\ \hline
\multicolumn{6}{l}{{\normalsize $L^*=0$, $\mu^{*2}=9$}} \\ \hline
3.63830\dag   &   0.00513\0   (8) &   0.87156    (21) &   0.00145\0   (2) & -40.872\0\0  (11) &  -0.3687\0   (60) \\
5.29200       &   0.1199\0   (17) &   0.69981    (55) &   0.02821    (48) & -34.055\0\0  (20) &  -3.475\0\0  (63) \\
6.28430       &   0.3909\0   (35) &   0.5362\0   (19) &   0.1082\0   (30) & -27.867\0\0  (60) &  -9.37\0\0\0 (22) \\ \hline
\multicolumn{6}{l}{{\normalsize $L^*=0$, $\mu^{*2}=12$}} \\ \hline
4.01780       &   0.00435\0  (91) &   0.87717    (26) &   0.00111\0  (24) & -46.773\0\0  (12) &  -0.483\0\0 (104) \\
5.84400       &   0.1191\0   (19) &   0.69862    (53) &   0.02602    (39) & -38.948\0\0  (20) &  -4.482\0\0  (44) \\
6.93975       &   0.3852\0   (45) &   0.5301\0   (20) &   0.0889\0   (62) & -32.038\0\0  (67) & -10.51\0\0\0 (53) \\ \hline
\multicolumn{6}{l}{{\normalsize $L^*=0$, $\mu^{*2}=16$}} \\ \hline
4.53370       &   0.00323    (35) &   0.88692    (27) &   0.00073\0   (8) & -55.163\0\0  (19) &  -0.626\0\0  (70) \\
6.77600       &   0.1352\0   (28) &   0.67890    (68) &   0.02720    (96) & -45.066\0\0  (27) &  -6.84\0\0\0 (29) \\
7.91330       &   0.4209\0   (54) &   0.5062\0   (31) &   0.1000\0   (26) & -37.29\0\0\0 (11) & -15.05\0\0\0 (35) \\ \hline
\multicolumn{6}{l}{{\normalsize $L^*=0$, $\mu^{*2}=20$}} \\ \hline
5.11670       &   0.00305    (48) &   0.89663    (41) &   0.00062    (10) & -64.216\0\0  (33) &  -0.94\0\0\0 (15) \\
7.60000       &   0.1335\0   (33) &   0.67349    (71) &   0.0248\0   (11) & -52.230\0\0  (28) &  -9.14\0\0\0 (44) \\
8.93080       &   0.4475\0   (92) &   0.4862\0   (89) &   0.1029\0   (41) & -42.93\0\0\0 (32) & -19.56\0\0\0 (62) \\ \hline
\multicolumn{6}{l}{{\normalsize $L^*=0.2$, $\mu^{*2}=0$}} \\ \hline
2.42000\dag   &   0.00576    (31) &   0.73095    (24) &   0.00244    (13) & -22.5710\0   (79) &  -0.1737\0   (96) \\
3.52000       &   0.1092\0   (11) &   0.58419    (52) &   0.03850    (56) & -18.792\0\0  (14) &  -2.024\0\0  (47) \\
4.18000       &   0.2990\0   (22) &   0.4236\0   (23) &   0.1279\0   (23) & -14.595\0\0  (52) &  -5.769\0\0  (94) \\ \hline
\end{tabular}}
\end{center}
\end{table}

\begin{table}[ht]
\noindent
\begin{center}
Table \ref{t2cljdvle}: continued. \\
\end{center}\medskip
\begin{center}
{\footnotesize
\begin{tabular}{l|ccccc}\hline\hline
{\normalsize ~$\!T^*$} &
{\normalsize $p_{\sigma}^*$} &
{\normalsize $\rho'^*$} & {\normalsize $\rho''^*$} &
{\normalsize $h'^{\mbox{\scriptsize res*}}$} &
{\normalsize $h''^{\mbox{\scriptsize res*}}$} \\ \hline
\multicolumn{6}{l}{{\normalsize $L^*=0.2$, $\mu^{*2}=3$}} \\ \hline
2.53720\dag   &   0.00506\0   (6) &   0.74233    (15) &   0.00204\0   (2) & -25.1959\0   (76) &  -0.1784\0   (21) \\
3.69040       &   0.1080\0   (10) &   0.59307    (52) &   0.03727    (31) & -20.711\0\0  (15) &  -2.292\0\0  (19) \\
4.38240       &   0.3046\0   (26) &   0.4289\0   (35) &   0.1265\0   (32) & -15.990\0\0  (86) &  -6.34\0\0\0 (15) \\ \hline
\multicolumn{6}{l}{{\normalsize $L^*=0.2$, $\mu^{*2}=6$}} \\ \hline
2.74340\dag   &   0.00397\0   (5) &   0.75312    (20) &   0.00148\0   (2) & -29.2578\0   (80) &  -0.2074\0   (26) \\
3.99040       &   0.0971\0   (10) &   0.60338    (46) &   0.03053    (47) & -24.126\0\0  (15) &  -2.555\0\0  (45) \\
4.73860       &   0.2965\0   (27) &   0.4490\0   (24) &   0.1115\0   (30) & -19.121\0\0  (64) &  -6.97\0\0\0 (18) \\ \hline
\multicolumn{6}{l}{{\normalsize $L^*=0.2$, $\mu^{*2}=9$}} \\ \hline
2.98490\dag   &   0.00315\0   (5) &   0.76127    (17) &   0.00108\0   (2) & -33.6769\0   (83) &  -0.2596\0   (40) \\
4.34160       &   0.0845\0   (11) &   0.60947    (45) &   0.02362    (29) & -27.941\0\0  (15) &  -2.726\0\0  (38) \\
5.15570       &   0.2825\0   (29) &   0.4630\0   (18) &   0.1001\0   (30) & -22.710\0\0  (53) &  -8.23\0\0\0 (20) \\ \hline
\multicolumn{6}{l}{{\normalsize $L^*=0.2$, $\mu^{*2}=12$}} \\ \hline
3.28790\dag   &   0.00319\0   (5) &   0.76489    (21) &   0.00100\0   (2) & -38.165\0\0  (11) &  -0.3912\0   (59) \\
4.78240       &   0.0830\0   (17) &   0.60746    (47) &   0.0229\0   (10) & -31.755\0\0  (17) &  -3.84\0\0\0 (20) \\
5.67910       &   0.2921\0   (31) &   0.4581\0   (16) &   0.0966\0   (44) & -25.958\0\0  (48) &  -9.82\0\0\0 (36) \\ \hline
\multicolumn{6}{l}{{\normalsize $L^*=0.2$, $\mu^{*2}=16$}} \\ \hline
3.71250       &   0.00191    (57) &   0.76778    (17) &   0.00053    (16) & -44.4425\0   (75) &  -0.37\0\0\0 (11) \\
5.44000       &   0.0913\0   (20) &   0.59958    (58) &   0.02294    (83) & -36.866\0\0  (22) &  -5.47\0\0\0 (24) \\
6.41250       &   0.3067\0   (36) &   0.4424\0   (22) &   0.0966\0   (26) & -30.213\0\0  (73) & -12.92\0\0\0 (29) \\ \hline
\multicolumn{6}{l}{{\normalsize $L^*=0.4$, $\mu^{*2}=0$}} \\ \hline
1.78750\dag   &   0.00326\0   (3) &   0.59571    (11) &   0.00187\0   (2) & -17.2882\0   (60) &  -0.1314\0   (14) \\
2.60000       &   0.06524    (64) &   0.47244    (58) &   0.03277    (80) & -14.213\0\0  (14) &  -1.708\0\0  (73) \\
3.08750       &   0.1892\0   (18) &   0.3417\0   (26) &   0.1231\0   (59) & -10.955\0\0  (54) &  -5.08\0\0\0 (21) \\ \hline
\multicolumn{6}{l}{{\normalsize $L^*=0.4$, $\mu^{*2}=3$}} \\ \hline
1.86835\dag   &   0.00271\0   (3) &   0.60327    (11) &   0.00149\0   (2) & -19.2066\0   (54) &  -0.1316\0   (16) \\
2.71760       &   0.06292    (78) &   0.47941    (49) &   0.0317\0   (13) & -15.655\0\0  (13) &  -1.97\0\0\0 (13) \\
3.22715       &   0.1830\0   (14) &   0.3362\0   (25) &   0.1048\0   (26) & -11.808\0\0  (58) &  -4.92\0\0\0 (12) \\ \hline
\multicolumn{6}{l}{{\normalsize $L^*=0.4$, $\mu^{*2}=6$}} \\ \hline
2.01025\dag   &   0.00205\0   (3) &   0.61042    (12) &   0.00104\0   (1) & -22.1685\0   (55) &  -0.1547\0   (20) \\
2.92400       &   0.05441    (67) &   0.48638    (43) &   0.02355    (39) & -18.139\0\0  (13) &  -1.993\0\0  (34) \\
3.43968       &   0.1632\0   (15) &   0.3710\0   (16) &   0.0839\0   (17) & -14.631\0\0  (42) &  -5.123\0\0  (89) \\ \hline
\multicolumn{6}{l}{{\normalsize $L^*=0.4$, $\mu^{*2}=9$}} \\ \hline
2.17415\dag   &   0.00156\0   (2) &   0.61655    (14) &   0.00073\0   (1) & -25.4563\0   (61) &  -0.1937\0   (29) \\
3.16240       &   0.04737    (81) &   0.49242    (42) &   0.01895    (42) & -20.979\0\0  (14) &  -2.271\0\0  (50) \\
3.75535       &   0.1605\0   (19) &   0.3697\0   (16) &   0.0770\0   (21) & -16.889\0\0  (46) &  -6.02\0\0\0 (13) \\ \hline
\end{tabular}}
\end{center}
\end{table}

\begin{table}[ht]
\noindent
\begin{center}
Table \ref{t2cljdvle}: continued. \\
\end{center}
\medskip
\begin{center}
{\footnotesize
\begin{tabular}{l|ccccc}\hline\hline
{\normalsize ~$\!T^*$} &
{\normalsize $p_{\sigma}^*$} &
{\normalsize $\rho'^*$} & {\normalsize $\rho''^*$} &
{\normalsize $h'^{\mbox{\scriptsize res*}}$} &
{\normalsize $h''^{\mbox{\scriptsize res*}}$} \\ \hline
\multicolumn{6}{l}{{\normalsize $L^*=0.4$, $\mu^{*2}=12$}} \\ \hline
2.38315\dag   &   0.00137\0   (2) &   0.61865    (16) &   0.00059\0   (1) & -28.7404\0   (84) &  -0.2600\0   (35) \\
3.46640       &   0.0455\0   (15) &   0.48990    (44) &   0.0179\0   (32) & -23.731\0\0  (15) &  -3.06\0\0\0 (69) \\
4.09920       &   0.1590\0   (28) &   0.3699\0   (16) &   0.0732\0   (33) & -19.401\0\0  (47) &  -7.23\0\0\0 (24) \\ \hline
\multicolumn{6}{l}{{\normalsize $L^*=0.505$, $\mu^{*2}=0$}} \\ \hline
1.55650\dag   &   0.00237\0   (3) &   0.54105    (10) &   0.00156\0   (2) & -15.3899\0   (52) &  -0.1103\0   (13) \\
2.26400       &   0.05086    (49) &   0.42814    (44) &   0.02815    (29) & -12.583\0\0  (11) &  -1.385\0\0  (17) \\
2.68850       &   0.1487\0   (15) &   0.2927\0   (50) &   0.1093\0   (38) &  -9.31\0\0\0 (10) &  -4.42\0\0\0 (14) \\ \hline
\multicolumn{6}{l}{{\normalsize $L^*=0.505$, $\mu^{*2}=3$}} \\ \hline
1.58070\dag   &   0.00147\0   (2) &   0.55324\0   (9) &   0.00095\0   (1) & -17.2624\0   (51) &  -0.0898\0   (12) \\
2.36500       &   0.04830    (58) &   0.43338    (43) &   0.02539    (69) & -13.835\0\0  (12) &  -1.464\0\0  (53) \\
2.80820       &   0.1479\0   (18) &   0.3121\0   (36) &   0.1016\0   (33) & -10.593\0\0  (82) &  -4.65\0\0\0 (15) \\ \hline
\multicolumn{6}{l}{{\normalsize $L^*=0.505$, $\mu^{*2}=6$}} \\ \hline
1.74520\dag   &   0.00148\0   (2) &   0.55403    (11) &   0.00087\0   (1) & -19.6812\0   (58) &  -0.1371\0   (19) \\
2.53800       &   0.04245    (64) &   0.44031    (36) &   0.02117    (49) & -16.031\0\0  (11) &  -1.758\0\0  (46) \\
3.01435       &   0.1365\0   (12) &   0.3203\0   (32) &   0.0849\0   (34) & -12.472\0\0  (75) &  -4.99\0\0\0 (17) \\ \hline
\multicolumn{6}{l}{{\normalsize $L^*=0.505$, $\mu^{*2}=9$}} \\ \hline
1.88320\dag   &   0.00108\0   (2) &   0.55953    (11) &   0.00059\0   (1) & -22.5779\0   (61) &  -0.1731\0   (27) \\
2.73900       &   0.03763    (74) &   0.44532    (42) &   0.0211\0   (23) & -18.511\0\0  (13) &  -2.77\0\0\0 (41) \\
3.25280       &   0.1252\0   (14) &   0.3321\0   (16) &   0.0715\0   (16) & -14.842\0\0  (41) &  -5.59\0\0\0 (11) \\ \hline
\multicolumn{6}{l}{{\normalsize $L^*=0.505$, $\mu^{*2}=12$}} \\ \hline
2.06030\dag   &   0.00096\0   (1) &   0.56200    (13) &   0.00048\0   (1) & -25.4946\0   (73) &  -0.2481\0   (35) \\
2.99000       &   0.0352\0   (10) &   0.44477    (41) &   0.0159\0   (30) & -20.985\0\0  (13) &  -2.73\0\0\0 (66) \\
3.56000       &   0.1256\0   (19) &   0.3223\0   (19) &   0.0702\0   (30) & -16.723\0\0  (56) &  -6.79\0\0\0 (20) \\ \hline
\multicolumn{6}{l}{{\normalsize $L^*=0.6$, $\mu^{*2}=0$}} \\ \hline
1.40250\dag   &   0.00197\0   (2) &   0.50039\0   (9) &   0.00144\0   (2) & -14.0889\0   (46) &  -0.1027\0   (11) \\
2.04000       &   0.04288    (48) &   0.39339    (41) &   0.02746    (70) & -11.408\0\0  (10) &  -1.387\0\0  (55) \\
2.42250       &   0.1286\0   (13) &   0.2643\0   (34) &   0.1075\0   (33) &  -8.299\0\0  (70) &  -4.26\0\0\0 (13) \\ \hline
\multicolumn{6}{l}{{\normalsize $L^*=0.6$, $\mu^{*2}=3$}} \\ \hline
1.40800\dag   &   0.00106\0   (2) &   0.51372\0   (8) &   0.00076\0   (1) & -15.8529\0   (52) &  -0.0751\0   (11) \\
2.12500       &   0.04056    (48) &   0.39966    (40) &   0.02425    (41) & -12.580\0\0  (11) &  -1.407\0\0  (25) \\
2.45952       &   0.1047\0   (11) &   0.3067\0   (13) &   0.0706\0   (38) & -10.124\0\0  (29) &  -3.35\0\0\0 (17) \\ \hline
\multicolumn{6}{l}{{\normalsize $L^*=0.6$, $\mu^{*2}=6$}} \\ \hline
1.56640\dag   &   0.00117\0   (2) &   0.51251    (11) &   0.00077\0   (1) & -18.0029\0   (61) &  -0.1286\0   (18) \\
2.27840       &   0.03499    (59) &   0.40589    (38) &   0.01938    (59) & -14.577\0\0  (11) &  -1.615\0\0  (54) \\
2.70560       &   0.1181\0   (14) &   0.2987\0   (32) &   0.0863\0   (27) & -11.399\0\0  (77) &  -4.91\0\0\0 (14) \\ \hline
\end{tabular}}
\end{center}
\end{table}
\clearpage

\begin{table}[ht]
\noindent
\begin{center}
Table \ref{t2cljdvle}: continued. \\
\end{center}
\medskip
\begin{center}
{\footnotesize
\begin{tabular}{l|ccccc}\hline\hline
{\normalsize ~$\!T^*$} &
{\normalsize $p_{\sigma}^*$} &
{\normalsize $\rho'^*$} & {\normalsize $\rho''^*$} &
{\normalsize $h'^{\mbox{\scriptsize res*}}$} &
{\normalsize $h''^{\mbox{\scriptsize res*}}$} \\ \hline
\multicolumn{6}{l}{{\normalsize $L^*=0.6$, $\mu^{*2}=9$}} \\ \hline
1.68800\dag   &   0.00082\0   (1) &   0.51756    (11) &   0.00050\0   (1) & -20.6624\0   (61) &  -0.1637\0   (23) \\
2.45500       &   0.0276\0   (12) &   0.40916    (39) &   0.01464    (57) & -16.807\0\0  (12) &  -1.888\0\0  (56) \\
2.88515       &   0.0979\0   (13) &   0.3160\0   (11) &   0.0629\0   (33) & -13.769\0\0  (27) &  -4.94\0\0\0 (21) \\ \hline
\multicolumn{6}{l}{{\normalsize $L^*=0.6$, $\mu^{*2}=12$}} \\ \hline
1.84250\dag   &   0.00071\0   (1) &   0.52036    (13) &   0.00040\0   (1) & -23.3629\0   (78) &  -0.2435\0   (31) \\
2.68000       &   0.0265\0   (15) &   0.40876    (41) &   0.0123\0   (13) & -19.059\0\0  (13) &  -2.06\0\0\0 (26) \\
3.15552       &   0.0989\0   (18) &   0.3071\0   (21) &   0.0632\0   (37) & -15.509\0\0  (56) &  -6.18\0\0\0 (26) \\ \hline
\multicolumn{6}{l}{{\normalsize $L^*=0.8$, $\mu^{*2}=0$}} \\ \hline
1.17700\dag   &   0.00135\0   (2) &   0.43871\0   (9) &   0.00118\0   (2) & -12.2841\0   (39) &  -0.0855\0   (11) \\
1.71200       &   0.03152    (36) &   0.34003    (50) &   0.02282    (22) &  -9.768\0\0  (12) &  -1.090\0\0  (12) \\
2.03300       &   0.0972\0   (11) &   0.2103\0   (44) &   0.0969\0   (32) &  -6.664\0\0  (91) &  -3.71\0\0\0 (11) \\ \hline
\multicolumn{6}{l}{{\normalsize $L^*=0.8$, $\mu^{*2}=3$}} \\ \hline
1.17150\dag   &   0.00065\0   (1) &   0.45198\0   (9) &   0.00057\0   (1) & -13.8571\0   (45) &  -0.05905    (90) \\
1.77200       &   0.02847    (34) &   0.34700    (35) &   0.02044    (37) & -10.8032\0   (93) &  -1.193\0\0  (26) \\
2.10430       &   0.0925\0   (10) &   0.2421\0   (96) &   0.0952\0   (36) &  -7.98\0\0\0 (20) &  -4.09\0\0\0 (13) \\ \hline
\multicolumn{6}{l}{{\normalsize $L^*=0.8$, $\mu^{*2}=6$}} \\ \hline
1.30130\dag   &   0.00070\0   (1) &   0.45104    (10) &   0.00055\0   (1) & -15.7904\0   (55) &  -0.1031\0   (15) \\
1.89280       &   0.02375    (53) &   0.35393    (41) &   0.01654    (77) & -12.575\0\0  (11) &  -1.482\0\0  (86) \\
2.24770       &   0.0846\0   (10) &   0.2567\0   (15) &   0.0687\0   (26) &  -9.692\0\0  (38) &  -4.12\0\0\0 (19) \\ \hline
\multicolumn{6}{l}{{\normalsize $L^*=0.8$, $\mu^{*2}=9$}} \\ \hline
1.38160\dag   &   0.00043\0   (1) &   0.45755    (10) &   0.00032\0   (1) & -18.3431\0   (57) &  -0.1412\0   (25) \\
2.02720       &   0.02041    (58) &   0.36185    (34) &   0.01305    (52) & -14.691\0\0  (10) &  -1.707\0\0  (65) \\
2.40730       &   0.0725\0   (11) &   0.2632\0   (21) &   0.0517\0   (19) & -11.524\0\0  (51) &  -4.07\0\0\0 (14) \\ \hline
\multicolumn{6}{l}{{\normalsize $L^*=0.8$, $\mu^{*2}=12$}} \\ \hline
1.52096\dag   &   0.00037\0   (1) &   0.46033    (13) &   0.00025\0   (1) & -20.8029\0   (69) &  -0.2326\0   (40) \\
2.19840       &   0.01671    (78) &   0.36363    (33) &   0.0084\0   (11) & -16.796\0\0  (11) &  -1.62\0\0\0 (23) \\
2.61060       &   0.0658\0   (12) &   0.2648\0   (12) &   0.0442\0   (44) & -13.318\0\0  (33) &  -4.62\0\0\0 (36) \\ \hline
\multicolumn{6}{l}{{\normalsize $L^*=1.0$, $\mu^{*2}=0$}} \\ \hline
0.97900\dag   &   0.00050\0   (1) &   0.41024\0   (7) &   0.00052\0   (1) & -11.5972\0   (43) &  -0.04220    (67) \\
1.50750       &   0.02504    (30) &   0.30729    (43) &   0.0184\0   (14) &  -8.871\0\0  (11) &  -0.931\0\0  (89) \\
1.69100       &   0.05413    (62) &   0.2461\0   (13) &   0.0436\0   (44) &  -7.358\0\0  (29) &  -1.93\0\0\0 (18) \\ \hline
\multicolumn{6}{l}{{\normalsize $L^*=1.0$, $\mu^{*2}=3$}} \\ \hline
1.02850\dag   &   0.00044\0   (1) &   0.41346\0   (9) &   0.00043\0   (1) & -12.8117\0   (48) &  -0.04662    (66) \\
1.50750       &   0.01711    (27) &   0.32581    (31) &   0.01378    (28) & -10.1499\0   (90) &  -0.848\0\0  (17) \\
1.77650       &   0.05688    (68) &   0.25218    (96) &   0.0495\0   (52) &  -8.122\0\0  (23) &  -2.39\0\0\0 (24) \\ \hline
\end{tabular}}
\end{center}
\end{table}
\clearpage

\begin{table}[ht]
\noindent
\begin{center}
Table \ref{t2cljdvle}: continued. \\
\end{center}
\medskip
\begin{center}
{\footnotesize
\begin{tabular}{l|ccccc}\hline\hline
{\normalsize ~$\!T^*$} &
{\normalsize $p_{\sigma}^*$} &
{\normalsize $\rho'^*$} & {\normalsize $\rho''^*$} &
{\normalsize $h'^{\mbox{\scriptsize res*}}$} &
{\normalsize $h''^{\mbox{\scriptsize res*}}$} \\ \hline
\multicolumn{6}{l}{{\normalsize $L^*=1.0$, $\mu^{*2}=6$}} \\ \hline
1.08900\dag   &   0.00026\0   (1) &   0.42225\0   (8) &   0.00024\0   (1) & -15.0377\0   (51) &  -0.0568\0   (11) \\
1.62000       &   0.01403    (48) &   0.32977    (34) &   0.01014    (65) & -11.761\0\0  (10) &  -0.921\0\0  (74) \\
1.90000       &   0.05034    (73) &   0.2575\0   (10) &   0.0464\0   (16) &  -9.533\0\0  (27) &  -2.98\0\0\0 (10) \\ \hline
\multicolumn{6}{l}{{\normalsize $L^*=1.0$, $\mu^{*2}=9$}} \\ \hline
1.17700\dag   &   0.00018\0   (1) &   0.42930\0   (9) &   0.00015\0   (1) & -17.5873\0   (90) &  -0.0976\0   (45) \\
1.71200       &   0.01020    (51) &   0.34335    (28) &   0.0103\0   (18) & -14.056\0\0  (10) &  -2.24\0\0\0 (56) \\
2.03300       &   0.04132    (86) &   0.26785    (87) &   0.0294\0   (28) & -11.437\0\0  (24) &  -2.77\0\0\0 (22) \\ \hline
\multicolumn{6}{l}{{\normalsize $L^*=1.0$, $\mu^{*2}=12$}} \\ \hline
1.30740\dag   &   0.00014\0   (1) &   0.43197    (11) &   0.00011\0   (1) & -20.166\0\0  (12) &  -0.1711\0   (40) \\
1.90160       &   0.01033    (65) &   0.34090    (33) &   0.00665    (27) & -15.996\0\0  (12) &  -1.484\0\0  (15) \\
2.28190       &   0.0501\0   (10) &   0.2482\0   (23) &   0.0471\0   (31) & -12.506\0\0  (62) &  -5.09\0\0\0 (32) \\ \hline
\end{tabular}}
\end{center}
\end{table}
\clearpage

\begin{table}[ht]
\noindent
\caption[]{Critical data, the critical compressibility factor, and the acentric factor of 38 2CLJD model fluids.}
\label{2CLJDcdomtab}
\bigskip
\begin{center}
\begin{tabular}{|c|c||l|l|l|l|l|l|l||l|}\hline
\multicolumn{2}{|c||}{ } & \multicolumn{7}{|l||}{$L^*$} & \\ \cline{3-9}
\multicolumn{2}{|c||}{ } & $0$ & $0.2$ & $0.4$ & $0.505$ & $0.6$ & $0.8$ & $1.0$ & \\ \hline \hline
$\mu^{*2}$ & 0 & \phantom{-}5.236  & \phantom{-}4.313  & \phantom{-}3.163  & \phantom{-}2.735  & \phantom{-}2.454  & \phantom{-}2.049  & \phantom{-}1.762  & $T^*_{\rm c}$ \\ \cline{3-10}
           &   & \phantom{-}0.3143 & \phantom{-}0.2740 & \phantom{-}0.2251 & \phantom{-}0.2032 & \phantom{-}0.1850 & \phantom{-}0.1577 & \phantom{-}0.1453 & $\rho^*_{\rm c}$ \\ \cline{3-10}
           &   & \phantom{-}0.4736 & \phantom{-}0.3670 & \phantom{-}0.2116 & \phantom{-}0.1616 & \phantom{-}0.1353 & \phantom{-}0.0995 & \phantom{-}0.0794 & $p^*_{\rm c}$ \\ \cline{3-10}
           &   & \phantom{-}0.2878 & \phantom{-}0.3106 & \phantom{-}0.2972 & \phantom{-}0.2908 & \phantom{-}0.2980 & \phantom{-}0.3079 & \phantom{-}0.3101 & $Z_{\rm c}$ \\ \cline{3-10}
           &   &           -0.0542 &           -0.0245 & \phantom{-}0.0308 & \phantom{-}0.0514 & \phantom{-}0.0589 & \phantom{-}0.0755 & \phantom{-}0.1382 & $\omega$ \\ \cline{2-10}

           & 3 & \phantom{-}5.475  & \phantom{-}4.521  & \phantom{-}3.311  & \phantom{-}2.871  & \phantom{-}2.558  & \phantom{-}2.127  & \phantom{-}1.876  & $T^*_{\rm c}$ \\ \cline{3-10}
           &   & \phantom{-}0.3197 & \phantom{-}0.2808 & \phantom{-}0.2237 & \phantom{-}0.2028 & \phantom{-}0.1868 & \phantom{-}0.1611 & \phantom{-}0.1449 & $\rho^*_{\rm c}$ \\ \cline{3-10}
           &   & \phantom{-}0.5157 & \phantom{-}0.3619 & \phantom{-}0.2133 & \phantom{-}0.1711 & \phantom{-}0.1378 & \phantom{-}0.0932 & \phantom{-}0.0856 & $p^*_{\rm c}$ \\ \cline{3-10}
           &   & \phantom{-}0.2946 & \phantom{-}0.2851 & \phantom{-}0.2880 & \phantom{-}0.2939 & \phantom{-}0.2884 & \phantom{-}0.2720 & \phantom{-}0.3149 & $Z_{\rm c}$ \\ \cline{3-10}
           &   &           -0.0191 & \phantom{-}0.0055 & \phantom{-}0.0528 & \phantom{-}0.0720 & \phantom{-}0.0958 & \phantom{-}0.1475 & \phantom{-}0.1797 & $\omega$ \\ \cline{2-10}

           & 6 & \phantom{-}5.990  & \phantom{-}4.917  & \phantom{-}3.591  & \phantom{-}3.100  & \phantom{-}2.774  & \phantom{-}2.316  & \phantom{-}2.010  & $T^*_{\rm c}$\\ \cline{3-10}
           &   & \phantom{-}0.3169 & \phantom{-}0.2781 & \phantom{-}0.2202 & \phantom{-}0.2017 & \phantom{-}0.1839 & \phantom{-}0.1576 & \phantom{-}0.1428 & $\rho^*_{\rm c}$ \\ \cline{3-10}
           &   & \phantom{-}0.5179 & \phantom{-}0.3624 & \phantom{-}0.2105 & \phantom{-}0.1677 & \phantom{-}0.1342 & \phantom{-}0.0894 & \phantom{-}0.0824 & $p^*_{\rm c}$ \\ \cline{3-10}
           &   & \phantom{-}0.2728 & \phantom{-}0.2650 & \phantom{-}0.2662 & \phantom{-}0.2682 & \phantom{-}0.2631 & \phantom{-}0.2449 & \phantom{-}0.2871 & $Z_{\rm c}$ \\ \cline{3-10}
           &   & \phantom{-}0.0214 & \phantom{-}0.0471 & \phantom{-}0.0976 & \phantom{-}0.1179 & \phantom{-}0.1429 & \phantom{-}0.1987 & \phantom{-}0.2398 & $\omega$ \\ \cline{2-10}

           & 9 & \phantom{-}6.585  & \phantom{-}5.382  & \phantom{-}3.909  & \phantom{-}3.377  & \phantom{-}3.040  & \phantom{-}2.520  & \phantom{-}2.178  & $T^*_{\rm c}$ \\ \cline{3-10}
           &   & \phantom{-}0.3124 & \phantom{-}0.2700 & \phantom{-}0.2164 & \phantom{-}0.1959 & \phantom{-}0.1790 & \phantom{-}0.1510 & \phantom{-}0.1382 & $\rho^*_{\rm c}$ \\ \cline{3-10}
           &   & \phantom{-}0.5098 & \phantom{-}0.3594 & \phantom{-}0.2078 & \phantom{-}0.1646 & \phantom{-}0.1304 & \phantom{-}0.0839 & \phantom{-}0.0765 & $p^*_{\rm c}$ \\ \cline{3-10}
           &   & \phantom{-}0.2478 & \phantom{-}0.2473 & \phantom{-}0.2457 & \phantom{-}0.2488 & \phantom{-}0.2396 & \phantom{-}0.2205 & \phantom{-}0.2542 & $Z_{\rm c}$ \\ \cline{3-10}
           &   & \phantom{-}0.0687 & \phantom{-}0.0938 & \phantom{-}0.1457 & \phantom{-}0.1669 & \phantom{-}0.1936 & \phantom{-}0.2565 & \phantom{-}0.3114 & $\omega$ \\ \cline{2-10}

           &12 & \phantom{-}7.289  & \phantom{-}5.945  & \phantom{-}4.282  & \phantom{-}3.676  & \phantom{-}3.273  & \phantom{-}2.713  & \phantom{-}2.370  & $T^*_{\rm c}$ \\ \cline{3-10}
           &   & \phantom{-}0.3033 & \phantom{-}0.2647 & \phantom{-}0.2122 & \phantom{-}0.1921 & \phantom{-}0.1765 & \phantom{-}0.1519 & \phantom{-}0.1376 & $\rho^*_{\rm c}$ \\ \cline{3-10}
           &   & \phantom{-}0.5067 & \phantom{-}0.3619 & \phantom{-}0.2095 & \phantom{-}0.1651 & \phantom{-}0.1291 & \phantom{-}0.0797 & \phantom{-}0.0716 & $p^*_{\rm c}$ \\ \cline{3-10}
           &   & \phantom{-}0.2292 & \phantom{-}0.2300 & \phantom{-}0.2306 & \phantom{-}0.2338 & \phantom{-}0.2235 & \phantom{-}0.1934 & \phantom{-}0.2196 & $Z_{\rm c}$ \\ \cline{3-10}
           &   & \phantom{-}0.1159 & \phantom{-}0.1396 & \phantom{-}0.1921 & \phantom{-}0.2145 & \phantom{-}0.2435 & \phantom{-}0.3163 & \phantom{-}0.3881 & $\omega$ \\ \cline{1-10}
\end{tabular}
\end{center}
\end{table}
\clearpage

\begin{table}[ht]
\noindent
\begin{center}
Table \ref{2CLJDcdomtab}: continued. \\
\end{center}
\medskip
\begin{center}
\begin{tabular}{|c|c||l|l|l|l|l|l|l||l|}\hline
\multicolumn{2}{|c||}{ } & \multicolumn{7}{|l||}{$L^*$} & \\ \cline{3-9}
\multicolumn{2}{|c||}{ } & $0$ & $0.2$ & $0.4$ & $0.505$ & $0.6$ & $0.8$ & $1.0$ & \\ \hline \hline
$\mu^{*2}$ &16 & \phantom{-}8.249  & \phantom{-}6.654  &                   &                   &                   &                   &                   & $T^*_{\rm c}$ \\ \cline{3-10}
           &   & \phantom{-}0.2961 & \phantom{-}0.2565 & \phantom{-0.0000} & \phantom{-0.0000} & \phantom{-0.0000} & \phantom{-0.0000} & \phantom{-0.0000} & $\rho^*_{\rm c}$ \\ \cline{3-10}
           &   & \phantom{-}0.5202 & \phantom{-}0.3795 &                   &                   &                   &                   &                   & $p^*_{\rm c}$ \\ \cline{3-10}
           &   & \phantom{-}0.2130 & \phantom{-}0.2224 &                   &                   &                   &                   &                   & $Z_{\rm c}$ \\ \cline{3-10}
           &   & \phantom{-}0.1732 & \phantom{-}0.1945 &                   &                   &                   &                   &                   & $\omega$ \\ \cline{2-10}

           &20 & \phantom{-}9.164  &                   &                   &                   &                   &                   &                   & $T^*_{\rm c}$ \\ \cline{3-10}
           &   & \phantom{-}0.2884 &                   &                   &                   &                   &                   &                   & $\rho^*_{\rm c}$ \\ \cline{3-10}
           &   & \phantom{-}0.5567 &                   &                   &                   &                   &                   &                   & $p^*_{\rm c}$ \\ \cline{3-10}
           &   & \phantom{-}0.2106 &                   &                   &                   &                   &                   &                   & $Z_{\rm c}$ \\ \cline{3-10}
           &   & \phantom{-}0.2220 &                   &                   &                   &                   &                   &                   & $\omega$ \\ \hline
\end{tabular}
\end{center}
\end{table}
\clearpage

\listoffigures
\clearpage

\clearpage

\clearpage

\clearpage

\noindent
{\bf \Large Appendix A}

\medskip
\noindent
{\bf \large Technical details}

The two-centre Lennard-Jones plus axial pointdipole (2CLJD) pair potential writes as 
\begin{eqnarray}
u_{\rm 2CLJD}(\bm{r}_{ij},\bm{\omega}_i,\bm{\omega}_j,L,\mu^2) = u_{\rm 2CLJ}(\bm{r}_{ij},\bm{\omega}_i,\bm{\omega}_j,L)+u_{\rm D}(\bm{r}_{ij},\bm{\omega}_i,\bm{\omega}_j,\mu^2), \nonumber
\end{eqnarray}
wherein $u_{\rm 2CLJ}$ is the Lennard-Jones part
\begin{eqnarray}
u_{\rm 2CLJ}(\bm{r}_{ij},\bm{\omega}_i,\bm{\omega}_j,L)=\sum_{a=1}^{2} \sum_{b=1}^{2} 4\epsilon \left[ \left( \frac{\sigma}{r_{ab}} \right)^{12} - \left( \frac{\sigma}{r_{ab}} \right)^6 \right], \nonumber
\end{eqnarray}
and $u_{\rm D}$ is the dipolar part, as given by Gray and Gubbins \cite{gray84}
\begin{eqnarray}
u_{\rm D}(\bm{r}_{ij},\bm{\omega}_i,\bm{\omega}_j,\mu^2)=\frac{1}{4 \pi \epsilon_0} \cdot \frac{\mu^2}{\left|\bm{r}_{ij}\right|^3} \left( {\rm cos} \gamma_{ij} - 3 {\rm cos} \theta_i {\rm cos} \theta_j  \right).
\label{uD}
\end{eqnarray}
Herein ${\bm r}_{ij}$ is the centre-centre distance vector of two molecules $i$ and $j$, $r_{ab}$ is one of the four Lennard-Jones site-site distances; $a$ counts the two sites of molecule $i$, $b$ counts those of molecule $j$. The vectors ${\bm \omega}_i$ and ${\bm \omega}_j$ represent the orientations of the two molecules $i$ and $j$. $\theta_i$ and $\theta_j$ are the angles between the dipole vectors ${\bm \mu}_i$ and ${\bm \mu}_j$  of the molecules $i$ and $j$ respectively and their centre-centre distance vector ${\bm r}_{ij}$. The cosine of the angle $\gamma_{ij}$ between the dipole vectors ${\bm \mu}_i$ and ${\bm \mu}_j$ is calculated as ${\rm cos} \gamma_{ij} = \left( {\bm \mu}_i \cdot {\bm \mu}_j \right) / \left( \left|{\bm \mu}_i \right| \cdot \left|{\bm \mu}_j \right| \right)$. The number of parameters related to the dipole is reduced to one, namely the dipolar momentum $\mu$, as its position and orientation within the molecule are fixed and as it is reduced by the large distance approximation to a pointdipole whose interaction is described by Eq. (\ref{uD}). The Lennard-Jones parameters $\sigma$ and $\epsilon$ represent size and energy respectively.

Beyond a certain elongation $L$, for small intermolecular distances $|\bm{r}_{ij}|$ the $u_{\rm 2CLJD}$ pair potential might diverge to infinity, as the positive Len\-nard-Jones term $u_{\rm 2CLJ}$ cannot outweigh the divergence to infinity of the dipolar term $u_{\rm D}$, that occurs for some relative orientations of the molecules $i$ and $j$. This divergence leads to infinite Boltzmann factors, i.e. non-existence of the configurational integral. During phase space sampling by molecular simulation within the pressure range in question, this artefact of the 2CLJD potential causes no problem as intermolecular centre-centre distances are very improbable to fall below critical values. However, the calculation of the chemical potential by test particle insertion often runs into critical intermolecular centre-centre distances. To avoid computational problems in such cases, the dipole site was shielded by a hard sphere of diameter $0.4 \cdot \sigma$. This shielding of the dipole site is in analogy to the suggestion of M\"oller and Fischer \cite{moeller9435} to shield the quadrupole site in two-centre Lennard-Jones plus pointquadrupole fluids, where similar problems occur. This hard sphere was not active during configuration generation. For reasons of consistency, this shielding by a hard sphere with diameter $0.4 \cdot \sigma$ was applied to all fluids studied here.

In order to achieve a monotonous transition from $L>0$ to $L=0$ the spherical fluids 1CLJ, where $\mu^2=0$, and 1CLJD (Stockmayer fluid), where $L>0$ were treated as two-centre LJ fluids with $L=0$. This leads to site superposition that is not present when 1CLJ and 1CLJD fluids are represented by a simple LJ site. Therefore the reduced temperatures, reduced pressures, reduced enthalpies and reduced dipolar momenta here are fourfold of the corresponding values in the one LJ site case. Densities, of course, are not concerned.

For all simulations, the centre-centre cut-off radius $r_{\rm c}$ was set to $5.0 \cdot \sigma$. Outside the cut-off sphere the fluid was assumed to have no preferential relative orientations of the molecules, i.e., in the calculation of the LJ long range corrections for the potential internal energy, the virial, and the chemical potential, orientational averaging was done with equally weighted relative orientations as proposed by Lustig \cite{lustig8817}. Long distance corrections for the dipolar part of the potential model were calculated with the reaction field method \cite{barker7378,saager9127}. The reaction field is derived from the polarization of the dielectric continuum supposed to surround the cut-off sphere of molecule $i$ and is calculated as
\begin{eqnarray}
{\bm E}_{{\rm RF,}i} = \frac{2 \left( \epsilon_{\rm s} - 1 \right)}{2 \epsilon_{\rm s} + 1} \cdot \frac{1}{r_{\rm c}^3} \cdot \sum_{\stackrel{j=1}{r_{ij}<r_{\rm c}}}^{N} {\bm \mu}_j. \nonumber
\end{eqnarray}
For sufficiently large systems, i.e. $N \ge 500$, the sensitivity of simulation results to the value of the relative permittivity $\epsilon_{\rm s}$ is negligible, cf. Saager et al. \cite{saager9127}. Therefore, the relative permittivity $\epsilon_{\rm s}$ is usually set to infinity in simulations of dipolar fluids, cf. \cite{lisal9719,garzon9436,jedlovszky9521}. For reasons of consistency, in this work, the relative permittivity $\epsilon_{\rm s}$ was uniformly set to infinity.

The interaction of the reaction field ${\bm E}_{{\rm RF,}i}$ with the dipole $\mu_i$ of molecule $i$ contributes to the potential energy of this molecule
\begin{eqnarray}
u_{{\rm RF,}i} = - {\bm \mu}_i \cdot {\bm E}_{{\rm RF,}i}. \nonumber
\end{eqnarray}
The dipolar potential internal energy per particle of the system of $N$ 2CLJD particles is then
\begin{eqnarray}
u_{\rm D}^{\rm tot} = \frac{1}{N} \sum_{i=1}^{N-1} \sum_{\stackrel{j=i+1}{r_{ij}<r_{\rm c}}}^{N} \left[ u_{\rm D}(\bm{r}_{ij},\bm{\omega}_i,\bm{\omega}_j,\mu^2) - \frac{1}{4 \pi \epsilon_0} \cdot {\bm \mu}_i \cdot \frac{2 \left( \epsilon_{\rm s} -1 \right)}{2 \epsilon_{\rm s} +1} \cdot \frac{1}{r_{\rm c}^3} \cdot {\bm \mu}_j \right] - \frac{1}{4 \pi \epsilon_0} \cdot \frac{\epsilon_{\rm s}-1}{2 \epsilon_{\rm s}+1} \cdot \frac{1}{r_{\rm c}^3} \cdot \mu^2. \nonumber
\end{eqnarray}
Knowing $u_{\rm D}^{\rm tot}$, the dipolar contribution to the virial is easily calculated by
\begin{eqnarray}
w_{\rm D}^{\rm tot} = -3 \cdot u_{\rm D}^{\rm tot}. \nonumber
\end{eqnarray}
The homogeneous reaction field ${\bm E}_{{\rm RF,}i}$ in each cut-off sphere exerts no force on the dipole ${\bm \mu}_i$ in the centre of the cut-off sphere, however, its contribution to the torque on molecule $i$ is
\begin{eqnarray}
{\bm \tau}_{{\rm RF,}i} = {\bm \mu}_i \times {\bm E}_{{\rm RF,}i}, \nonumber
\end{eqnarray}
which has to be added when molecular dynamics simulations are considered.
\clearpage

\noindent
{\bf \Large Appendix B}

\medskip
\noindent
{\bf \large Correlation of critical properties}

The development of correlations in the present work was done in analogy to a previous work on the 2CLJQ fluid \cite{stoll0133}. The functions $T^*_{\rm c}(\mu^{*2},L^*)$ and $\rho^*_{\rm c}(\mu^{*2},L^*)$ were assumed to be linear combinations of elementary functions one of which is a constant $c$, the others depend either on $\mu^{*2}$, i.e. $\psi_i \left( \mu^{*2} \right)$, or on $L^*$, i.e. $\xi_i \left( L^* \right)$, or on both, i.e. $\chi_i \left( \mu^{*2},L^* \right)$. The number of elementary functions was restricted to up to two for both the $\mu^{*2}$- and the $L^*$-dependence and to up to three mixed terms. With $y$ representing any of the aforementioned functions, the linear combination writes as
\begin{eqnarray}
y \left( \mu^{*2},L^* \right) = c + \sum_{i=1}^{\le 2}\alpha_i \cdot \psi_i(\mu^{*2}) + \sum_{j=1}^{\le 2}\beta_j \cdot \xi_j(L^*) + \sum_{k=1}^{\le 3} \gamma_k \cdot \chi_k(\mu^{*2},L^*).
\label{lincomb}
\end{eqnarray}
Usual non-weighted least squares fits of these linear combinations to the critical data given in Table \ref{2CLJDcdomtab} yielded the coefficients in Table \ref{corrtab}, which also contains the elementary functions.
It should be mentioned, that the elementary function $\xi_i \left( L^* \right)$ from \cite{stoll0133} could be reused here, whereas new elementary functions $\psi_i$ and $\chi_i$ were selected due to different macroscopic thermodynamics of dipolar and quadrupolar fluids.

\medskip
\noindent
{\bf \large Correlation of saturated densities and vapour pressures}

The saturated density -- temperature correlations are based on Eqs. (\ref{Trho1corrcore}) and (\ref{Trho2corrcore}). The dependence on $\mu^{*2}$ and $L^*$ was ascribed to $T^*_{\rm c}$, $\rho^*_{\rm c}$, and to the coefficients $C_1$ to $C''_3$. The coefficient functions $C_1(\mu^{*2},L^*)$ to $C''_3(\mu^{*2},L^*)$ were linear combinations of elementary functions in the sense of Eq. (\ref{lincomb}). The correlations $T^*_{\rm c} \left( \mu^{*2},L^* \right)$ and $\rho^*_{\rm c}\left( \mu^{*2},L^* \right)$ were used in the fit of the functions $\rho'^*(\mu^{*2},L^*,T^*)$ and $\rho''^*(\mu^{*2},L^*,T^*)$ to data from simulation. The set of elementary functions used in the coefficient functions $C_1(\mu^{*2},L^*)$ to $C''_3(\mu^{*2},L^*)$ were different from those used for the 2CLJQ fluid \cite{stoll0133}.

For the vapour pressure -- temperature correlation the polynomial ansatz
\begin{eqnarray}
\ln p^*_{\sigma}(\mu^{*2},L^*,T^*)=c_1(\mu^{*2},L^*)+\frac{c_2(\mu^{*2},L^*)}{T^*}+\frac{c_3(\mu^{*2},L^*)}{T^{*4}},
\label{pTcorr}
\end{eqnarray}
was applied in analogy to \cite{stoll0133}. The coefficients $c_1(\mu^{*2},L^*)$ to $c_3(\mu^{*2},L^*)$ were linear combinations of the elementary functions in the sense of Eq. (\ref{lincomb}). These elementary functions were different from those used for the 2CLJQ fluid \cite{stoll0133}.

The fit of the functions $\rho'^*(\mu^{*2},L^*,T^*)$, $\rho''^*(\mu^{*2},L^*,T^*)$, and $\ln p^*_{\sigma}(\mu^{*2},L^*,T^*)$ to data from simulation was performed by uncertainty-weighted least squares minimization of the functions $F$ and $G$ defined as
\begin{eqnarray}
F &=& \sum_i \left[ \phantom{+} \frac{1}{ \delta \rho'^{*2}_i} \left[ \rho'^* \left( \mu^{*2}_i,L^*_i,T^*_i \right) - \rho'^*_{{\rm sim},i} \right]^2 \right. \nonumber \\ \nonumber
&\phantom{=}& \phantom{\sum_i} + \left. \frac{1}{ \delta \rho''^{*2}_i} \left[ \rho''^* \left( \mu^{*2}_i,L^*_i,T^*_i \right) - \rho''^*_{{\rm sim},i} \right]^2 \right] \stackrel{\rm !}{=} {\rm min}, \\ \nonumber
G &=& \sum_i \frac{1}{ \left( \delta \ln p^*_{\sigma,i} \right)^2} \left[ \ln p^*_{\sigma} \left( \mu^{*2}_i,L^*_i,T^*_i \right) - \ln p^*_{\sigma,{\rm sim},i} \right]^2 \stackrel{\rm !}{=} {\rm min}. \nonumber
\end{eqnarray}
The resulting coefficients and the elementary functions are given in Table \ref{corrtab}.

\clearpage
\begin{sidewaystable}[ht]
\noindent
\caption[]{Elementary functions and their coefficients for the correlations $T^*_{\rm c} \left( \mu^{*2},L^* \right)$, $\rho^*_{\rm c} \left( \mu^{*2},L^* \right)$, $C_1\left( \mu^{*2},L^* \right)$ to $C''_3\left( \mu^{*2},L^* \right)$, and $c_1\left( \mu^{*2},L^* \right)$ to $c_3\left( \mu^{*2},L^* \right)$. The notation is simplified: $\ell$ is $L^*$, $m$ is $\mu^{*2}$.}
\label{corrtab}
\bigskip
{\footnotesize
\begin{tabular}{|l|l|l|l|l|l|}\hline

\multicolumn{3}{|l|}{$T^*_{\rm c}\left(m,\ell \right)$} & \multicolumn{3}{|l|}{$\rho^*_{\rm c}\left(m,\ell \right)$} \\ \hline
$c$      & 1                                   & $\phantom{+} 0.1454013 \cdot 10^1$    & $c$      & 1                                         & $\phantom{+} 0.3157828$            \\ \hline
$\psi_i$ & $m/(88+m)^2$                        & $\phantom{+} 0.1363894 \cdot 10^3$    & $\psi_i$ & $m/(88+m)^2$                              & $\phantom{+} 0.9871123 \cdot 10^1$ \\ \cline{2-3} \cline{5-6}
         & $m^2/(88+m)^3$                      & $\phantom{+} 0.2020243 \cdot 10^4$    &          & $m^2/(88+m)^3$                            & $-0.1461751 \cdot 10^3$            \\ \cline{1-3} \cline{4-6}
$\xi_i$  & $1/(0.1+\ell^2)$                    & $\phantom{+} 0.3269772$               & $\xi_i$  & $\ell^2/(0.11+\ell^2)$                    & $-0.1475616$                       \\ \cline{2-3} \cline{5-6}
         & $1/(0.1+\ell^5)$                    & $\phantom{+} 0.4910240 \cdot 10^{-1}$ &          & $\ell^5/(0.11+\ell^5)$                    & $-0.4152214 \cdot 10^{-1}$         \\ \cline{1-3} \cline{4-6}
$\chi_i$ & $m/((88+m)^2 \cdot (0.1+\ell^2))$   & $\phantom{+} 0.4239005 \cdot 10^2$    & $\chi_i$ & $m/(88+m)^2 \cdot \ell^2/(0.11+\ell^2)$   & $-0.1010584 \cdot 10^2$            \\ \cline{2-3} \cline{5-6}
         & $m^2/((88+m)^3 \cdot (0.1+\ell^2))$ & $\phantom{+} 0.6724083 \cdot 10^3$    &          & $m^2/(88+m)^3 \cdot \ell^2/(0.11+\ell^2)$ & $\phantom{+} 0.4105884 \cdot 10^2$ \\ \cline{2-3} \cline{5-6}
         & $m^2/((88+m)^3 \cdot (0.1+\ell^5))$ & $\phantom{+} 0.7913876 \cdot 10^2$    &          & $m^2/(88+m)^3 \cdot \ell^5/(0.11+\ell^5)$ & $\phantom{+} 0.5299302 \cdot 10^2$ \\ \cline{1-3} \cline{4-6}

\multicolumn{3}{|l|}{$C_1\left(m,\ell \right)$} & \multicolumn{3}{|l|}{$C'_2\left(m,\ell \right)$} \\ \hline
$c$      & 1                                        & $\phantom{+} 0.2951644$            & $c$      & 1                                          & $\phantom{+} 0.6484789 \cdot 10^{-1}$ \\ \hline
$\psi_i$ & $m^2/(70+m)^2$                           & $-0.6339151$                       & $\psi_i$ & $m^2/(70+m)^2$                             & $\phantom{+} 0.7301440$               \\ \cline{2-3} \cline{5-6}
         & $m^3/(70+m)^3$                           & $\phantom{+} 0.3182745 \cdot 10^1$ &          & $m^3/(70+m)^3$                             & $-0.8780100 \cdot 10^1$               \\ \cline{1-3} \cline{4-6}
$\xi_i$  & $\ell^2 \cdot e^{\ell}$                  & $-0.2359527$                       & $\xi_i$  & $\ell^2 \cdot e^{\ell}$                    & $-0.6551324$                          \\ \cline{2-3} \cline{5-6}
         & $\ell^3$                                 & $\phantom{+} 0.5466755 $           &          & $\ell^3$                                   & $\phantom{+} 0.1810641 \cdot 10^1$    \\ \cline{1-3} \cline{4-6}
$\chi_i$ & $m^2/(70+m)^2 \cdot \ell^8/(\ell+0.4)$   & $\phantom{+} 0.1449170 \cdot 10^1$ & $\chi_i$ & $m^2/(70+m)^2 \cdot \ell^2 \cdot e^{\ell}$ & $\phantom{+} 0.4808117 \cdot 10^1$    \\ \cline{2-3} \cline{5-6}
         & $m^3/(70+m)^3 \cdot \ell^2/(\ell+0.4)^2$ & $-0.1955388$                       &          & $m^3/(70+m)^3 \cdot \ell^2 \cdot e^{\ell}$ & $\phantom{+} 0.1937455 \cdot 10^1$    \\ \cline{2-3} \cline{5-6}
         & $m^3/(70+m)^3 \cdot \ell^8/(\ell+0.4)$   & $-0.5849357 \cdot 10^1$            &          & $m^2/(70+m)^2 \cdot \ell^3$                & $-0.1320822 \cdot 10^2$               \\ \cline{1-3} \cline{4-6}

\multicolumn{3}{|l|}{$C'_3\left(m,\ell \right)$} & \multicolumn{3}{|l|}{$C''_2\left(m,\ell \right)$} \\ \hline
$c$      & 1                                        & $-0.7258204 \cdot 10^{-2}$        & $c$      & 1                                        & $-0.5486341 \cdot 10^{-2}$         \\ \hline
$\psi_i$ & $m^2/(70+m)^2$                           & $-0.6215183$                      & $\psi_i$ & $m^2/(70+m)^2$                           & $\phantom{+}0.1223952 \cdot 10^1$  \\ \cline{2-3} \cline{5-6}
         & $m^3/(70+m)^3$                           & $\phantom{+}0.4708560 \cdot 10^1$ &          & $m^3/(70+m)^3$                           & $\phantom{+}0.1350701 \cdot 10^1$  \\ \cline{1-3} \cline{4-6}
$\xi_i$  & $\ell^2 \cdot e^{\ell}$                  & $\phantom{+}0.4316296$            & $\xi_i$  & $\ell^2/(0.4+\ell)^2$                    & $\phantom{+}0.2479957$             \\ \cline{2-3} \cline{5-6}
         & $\ell^3$                                 & $-0.1166922 \cdot 10^1$           &          & $\ell^8/(0.4+\ell)$                      & $-0.1684560$                       \\ \cline{1-3} \cline{4-6}
$\chi_i$ & $m/(88+m)^2 \cdot \ell^2/(0.4+\ell)^2$   & $-0.2215021 \cdot 10^2$           & $\chi_i$ & $m/(88+m)^2 \cdot \ell^2/(0.4+\ell)^2$   & $-0.1956742 \cdot 10^2$            \\ \cline{2-3} \cline{5-6}
         & $m^2/(88+m)^3 \cdot \ell^2/(0.4+\ell)^2$ & $-0.7803181 \cdot 10^2$           &          & $m^2/(88+m)^3 \cdot \ell^2/(0.4+\ell)^2$ & $-0.2289032 \cdot 10^3$            \\ \cline{2-3} \cline{5-6}
         & $m/(88+m)^2 \cdot \ell^8/(0.4+\ell)$     & $-0.5735507$                      &          & $m^2/(88+m)^3 \cdot \ell^8/(0.4+\ell)$   & $\phantom{+}0.7221121 \cdot 10^3$  \\ \hline

\end{tabular}}
\end{sidewaystable}

\begin{sidewaystable}[ht]
\begin{center}
Table \ref{corrtab}: continued. \\
\end{center}
\medskip
{\footnotesize
\begin{tabular}{|l|l|l|l|l|l|}\hline

\multicolumn{3}{|l|}{$C''_3\left(m,\ell \right)$} & \multicolumn{3}{|l|}{$c_1\left(m,\ell \right)$} \\ \hline
$c$      & 1 & $\phantom{+} 0.2574709 \cdot 10^{-1}$ & $c$ & 1 & $\phantom{+} 0.4411718 \cdot 10^1$ \\ \hline
$\psi_i$ & $m/(88+m)^2$ & $-0.2940407 \cdot 10^1$ & $\psi_i$ & $m/(88+m)^2$ & $\phantom{+} 0.4575129 \cdot 10^3$ \\ \cline{2-3} \cline{5-6}
         & $m^2/(88+m)^3$ & $-0.1008706 \cdot 10^3$ & & $m^2/(88+m)^3$ & $\phantom{+} 0.2469929 \cdot 10^4$\\ \cline{1-3} \cline{4-6}
$\xi_i$  & $\ell^2/(0.4+\ell)^2$ & $-0.9426323 \cdot 10^{-1}$ & $\xi_i$ & $\ell^2/(0.4+\ell)^2$ & $-0.2016356 \cdot 10^1$ \\ \cline{2-3} \cline{5-6}
         & $\ell^8/(0.4+\ell)$  & $\phantom{+}0.1108324$ & & $\ell^8/(0.4+\ell)$ & $\phantom{+} 0.4346103$ \\ \cline{1-3} \cline{4-6}
$\chi_i$ & $m/(88+m)^2 \cdot \ell^2/(0.4+\ell)^2$ & $\phantom{+} 0.2543158 \cdot 10^2$ & $\chi_i$ & $m^2/(88+m)^3 \cdot \ell^2/(0.4+\ell)^2$ & $\phantom{+} 0.9787962 \cdot 10^3 $ \\ \cline{2-3} \cline{5-6}
         & $m^2/(88+m)^3 \cdot \ell^2/(0.4+\ell)^2$ & $\phantom{+} 0.5987224 \cdot 10^2$ & & $m^2/(88+m)^3 \cdot \ell^8/(0.4+\ell)$ & $\phantom{+} 0.2467171 \cdot 10^4$ \\ \cline{2-3} \cline{5-6}
         & $m^2/(88+m)^3 \cdot \ell^8/(0.4+\ell)$ & $-0.6511462 \cdot 10^3$ & & -- & -- \\ \cline{1-3} \cline{4-6}

\multicolumn{3}{|l|}{$c_2\left(m,\ell \right)$} & \multicolumn{3}{|l|}{$c_3\left(m,\ell \right)$} \\ \hline
$c$      & 1 & $-0.2686327 \cdot 10^2$ & $c$ & -- & -- \\ \hline
$\psi_i$ & $m/(88+m)^2$ & $-0.3428826 \cdot 10^4$ & $\psi_i$ & $m/(88+m)^2$ & $-0.5264689 \cdot 10^3$ \\ \cline{2-3} \cline{5-6}
         & $m^2/(88+m)^3$ & $-0.8720808 \cdot 10^5 $ & & $m^2/(88+m)^3$ & $\phantom{+} 0.6782756 \cdot 10^4$ \\ \cline{1-3} \cline{4-6}
$\xi_i$  & $\ell^2/(0.75+\ell)^2$ & $\phantom{+} 0.1275315 \cdot 10^3$ & $\xi_i$ & $\ell^4$ & $\phantom{+} 0.1812550 $ \\ \cline{2-3} \cline{5-6}
         & $\ell^3/(0.75+\ell)^3$ & $-0.1393077 \cdot 10^3$ & & -- & -- \\ \cline{1-3} \cline{4-6}
$\chi_i$ & $m/(88+m)^2 \cdot \ell^2/(0.75+\ell)^2$ & $\phantom{+} 0.7248855 \cdot 10^4$ & $\chi_i$ & -- & --  \\ \cline{2-3} \cline{5-6}
         & $m^2/(88+m)^3 \cdot \ell^2/(0.75+\ell)^2$ & $\phantom{+} 0.5715498 \cdot 10^6$ & & -- & -- \\ \cline{2-3} \cline{5-6}
         & $m^2/(88+m)^3 \cdot \ell^3/(0.75+\ell)^3$ & $-0.7433962 \cdot 10^6$ & & -- & -- \\ \cline{1-3} \cline{4-6}

\end{tabular}}
\end{sidewaystable}
\clearpage

\end{document}